\documentclass[%
reprint,
superscriptaddress,
amsmath,amssymb,
aps,longbibliography,
]{revtex4-1}

\usepackage{graphicx}
\usepackage{verbatim}
\usepackage[version=4]{mhchem}
\usepackage{xcolor}
\usepackage{physics}
\usepackage[normalem]{ulem}
\usepackage{pdfpages}
\usepackage{pgffor}
\usepackage{setspace}
\usepackage{times}

\usepackage{hyperref}
\usepackage[english]{babel}

\makeatletter
\AtBeginDocument{\let\LS@rot\@undefined}
\makeatother

\begin{document}

\setstretch{1.25}



\title{
Combining Simulations and Spectroscopy to Design Protective Coatings on Sulfide Solid Electrolytes}

\author{Zhiyi Wang}
\thanks{These authors contributed equally}
\affiliation{Laboratory of Computational Science and Modeling, Institut des Mat\'eriaux, \'Ecole Polytechnique F\'ed\'erale de Lausanne, 1015 Lausanne, Switzerland}

\author{Leonie Braks}
\thanks{These authors contributed equally}
\affiliation{Department of Chemistry, University of Fribourg, Chemin de Musee 9, Fribourg, 1700, Switzerland}

\author{Michele Ceriotti}

\affiliation{Laboratory of Computational Science and Modeling, Institut des Mat\'eriaux, \'Ecole Polytechnique F\'ed\'erale de Lausanne, 1015 Lausanne, Switzerland}

\author{Ali Coskun}

\affiliation{Department of Chemistry, University of Fribourg, Chemin de Musee 9, Fribourg, 1700, Switzerland}

\author{Hanna Türk}
\thanks{\seniorauthor}
\affiliation{Laboratory of Computational Science and Modeling, Institut des Mat\'eriaux, \'Ecole Polytechnique F\'ed\'erale de Lausanne, 1015 Lausanne, Switzerland}
\affiliation{Federal Institute for Materials Research and Testing, Berlin, Germany}

\author{Davide Tisi}
\thanks{\seniorauthor}
\affiliation{Laboratory of Computational Science and Modeling, Institut des Mat\'eriaux, \'Ecole Polytechnique F\'ed\'erale de Lausanne, 1015 Lausanne, Switzerland}
\affiliation{Department of Energy Conversion and Storage, Technical University of Denmark, Lyngby, Denmark}

\newcommand{\seniorauthor}{\protect\hypertarget{senior}{}These authors contributed equally to this work and share senior authorship. \href{mailto:hanna.tuerk@bam.de}{hanna.tuerk@bam.de}, \href{mailto:davti@dtu.dk}{davti@dtu.dk}}

\newcommand{\CSM}[1]{{\color{red}#1}}

\definecolor{Grey}{rgb}{0.50,0.50,0.50}
\definecolor{Blue}{rgb}{0.00,0.00,1.00}
\definecolor{Red}{rgb}{1.00,0.00,0.00}
\definecolor{Green}{rgb}{0.20,0.80,0.20}
\definecolor{Magenta}{rgb}{0.60,0.00,0.60}
\definecolor{BluBondi}{rgb}{0.00,0.58,0.71}
\definecolor{Orange}{rgb}{0.95,0.46,0.17}
\definecolor{tangerine}{rgb}{0.944,0.522,0}
\definecolor{violet}{rgb}{0.5059,0.3137,0.7098}
\newcommand{\editor}[2]{%
  \expandafter\newcommand\csname #1note\endcsname[1]{%
    \textcolor{#2}{{\it (\textbf{#1:} ##1)}}}%
  \expandafter\newcommand\csname #1\endcsname[1]{%
    \textcolor{#2}{##1}}%
  \expandafter\newcommand\csname #1cancel\endcsname[1]{%
    \textcolor{#2}{\sout{##1}}}%
  \expandafter\newcommand\csname #1change\endcsname[2]{%
    \textcolor{#2}{\sout{##1} ##2}}%
  \newenvironment{#1text}{\color{#2}}{\color{black}}
}
\editor{resub}{red}
\editor{DT}{tangerine}
\editor{HT}{Blue}
\editor{MC}{Green}
\editor{LB}{cyan}
\editor{ZW}{red}

\date{\today}

\begin{abstract}

Sulfide solid electrolytes are attractive for all-solid-state batteries (ASSBs) because of their high Li-ion conductivity and favorable processability, but their practical use is limited by electrochemical instability and moisture-induced degradation. Molecular surface coatings offer a scalable strategy to improve humidity tolerance and interfacial stability, yet there is limited understanding of the molecular-level factors governing surface interactions. By combining large-scale atomistic simulations enabled by a machine-learning potential with targeted experiments, we present a comprehensive framework for designing molecular coatings on Li$_6$PS$_5$Cl (LPSCl).
Using a PET-MAD interatomic potential fine-tuned on an r$^2$SCAN reference dataset covering LPSCl surfaces, molecular adsorbates, and their interactions, we model surface adsorption of carboxylic (CA), sulfonic (SA), and phosphonic (PA) acids with alkyl chains of varying lengths and the interaction of these coated surfaces with water at finite temperature.
The simulations show that lithiated head groups generally exhibit more favorable adsorption energies than their protonated counterparts. Fourier transform infrared (FTIR) spectra are consistent with in-situ lithiation of the adsorbed acids, suggesting that the lithiated form is the relevant surface species.
Furthermore, impedance spectroscopy shows that the lithiated coatings maintain 40--55\% of the initial ionic conductivity after powder processing. While adsorption strength is largely controlled by head-group chemistry and lithiation state, coating morphology depends primarily on alkyl-chain length and surface coverage. In particular, long-chain lithiated carboxylates form dense, brush-like hydrophobic layers that reduce water access to reactive Li surface sites, with protection setting in at a coating thickness of approximately 8~\AA{}. Consistent with these predictions, longer carboxylate chains suppress H$_2$S evolution more effectively. Because thicker coatings also hinder Li-ion transport, intermediate chain lengths offer the best balance between moisture protection and ionic conductivity.  Together, these results demonstrate how the interplay between interfacial binding and coating morphology governs the protection of sulfide electrolytes and provide guidelines for the rational design of interfaces in ASSBs.
\end{abstract}

\maketitle

\section{Introduction}
All-solid-state batteries (ASSBs) have attracted significant interest owing to their potential to deliver higher energy density and improved safety compared with state-of-the-art Li-ion batteries.~\cite{bates_are_2022, li_advance_2021, stegmaier_exploiting_2022,janek_challenges_2023,takada_progress_2018} In contrast to conventional Li-ion batteries, which rely on flammable liquid electrolytes, ASSBs employ solid electrolytes (SEs), thereby reducing safety risks associated with electrolyte leakage and flammability. Beyond safety, SEs can enable operation over wider temperature ranges,~\cite{kim_roles_2021, bates_are_2022}
 support faster charging by mitigating concentration polarization in the electrolyte,~\cite{janek_challenges_2023} and have the potential to enable higher energy and power densities through the use of Li-metal anodes and bipolar cell architectures.~\cite{jung_solid-state_2019}
 Over the past decades, several classes of SEs with high ionic conductivities have been developed, including oxide, polymer, halide, and sulfide solid electrolytes.~\cite{bouguern_comparative_2026,liu_deciphering_2025,tang_halide-based_2025} Among these, sulfide solid electrolytes (SSEs) are considered particularly promising because of their high room-temperature ionic conductivity and their good processability.
Within this family, argyrodite-based SSEs with the general formula of $\mathrm{Li_6PS_5X}$ (X = Cl, Br, I) have received considerable attention due to their ionic conductivities of up to 5 mS cm$^{-1}$ at room temperature. Despite these advantages, their adoption in industrially relevant applications remains limited.~\cite{forster_molecular_2025}
Besides the low intrinsic electrochemical stability of SSEs, commercialization is also hindered by the pronounced moisture sensitivity of SSEs, which leads to the release of toxic H$_2$S gas and degradation of the active material, even in dry rooms with low levels of water. This reactivity stems from the P–S framework, where the hard acid P readily reacts with the hard base O from atmospheric H$_2$O, replacing the softer sulfur atom. 
Various strategies have been explored to improve the moisture stability of SSEs, including the incorporation of softer acids to replace phosphorus and reduce susceptibility to attack by hard bases, as well as partial substitution of sulfur by oxygen. Recently, several groups have focused on coating SSEs with moisture-repellent layers,~\cite{Yu2022-C5Br,Kim2023-TMS,Liu2025-C11SH-moisture,Lee2025-CF3SH,kim_cascade_2026, Fadillah2026,qian_enabling_2026} in particular, octadecylphosphonic acid coatings proved to significantly improve the retention of ionic conductivity during dry-room exposure, while simultaneously improving electrochemical stability with both the anode and the cathode.~\cite{fadillah_molecular_2025}

Despite these efforts, the molecular-level principles governing coating effectiveness remain poorly understood. While a few studies include comparisons of different head groups,~\cite{Kim2023-TMS,Liu2025-C11SH-moisture} systematic investigations of head group chemistry are still lacking. Moreover, to the best of our knowledge, no studies have examined the influence of alkyl chain length. As a result, it is unclear which molecular features primarily control interfacial adhesion and which control moisture exclusion.~\cite{braks2026_review}

Addressing these challenges requires an atomistic-level understanding of the organic–inorganic interface, including surface reconstruction, head-group anchoring, the conformation and organization of the adsorbed molecular layer, and the structure and dynamics of interfacial water. 
Such detailed insight, however, is difficult to obtain experimentally owing to the limited number of characterization tools and remains challenging to achieve using conventional simulation methods. 
Recent experimental studies have demonstrated that molecular coatings can substantially improve the stability of sulfide electrolytes. In these studies,~\cite{qian_enabling_2026,shen_surfaceinterface_2021,marana_stability_2023,zhu_origin_2015} static DFT calculations provide valuable atomistic insight into coating adsorption and interfacial interactions. However, the computational cost of DFT limits simulations to small systems and short timescales, preventing the study of collective finite-temperature phenomena such as surface reconstruction, molecular self-assembly at realistic coverages, and water penetration. These processes require large-scale simulations extending over nanoseconds and involving thousands of atoms.

 The last decade saw the rise of machine-learning interatomic potentials (MLIPs),~\cite{zeng_deepmd-kit_2023,behler_generalized_2007,smith_ani-1_2017,bartok_gaussian_2010,bartok_representing_2013,rupp_fast_2012,butler_machine_2018,unke_spookynet_2021,batatia_design_2025,batatia_mace_2022,batzner_e3-equivariant_2022}
which hold the promise of overcoming this cost-accuracy trade-off. MLIPs, once trained on accurate quantum-mechanical reference data, reproduce DFT accuracy at a cost only marginally higher than empirical force fields.~\cite{miyagawa_accurate_2024,tisi_heat_2021,malosso_viscosity_2022,deringer_machine_2017,mazitov_surface_2024,lopanitsyna_finite-temperature_2021,lopanitsyna_modeling_2023,pegolo_temperature-_2022,sendek_machine_2022} MLIPs have already provided detailed mechanistic insight into the bulk and surface properties of lithium thiophosphates and related solid electrolytes, including their ionic and thermal transport properties.~\cite{gigli_mechanism_2024,tisi_thermal_2024,tuerk_reconstructions_2025,staacke_tackling_2022,staacke_role_2021,lee_disorder-dependent_2024,fragapane_lips_2026} 

\begin{figure}[t]
    \centering
    \includegraphics[width=\linewidth]{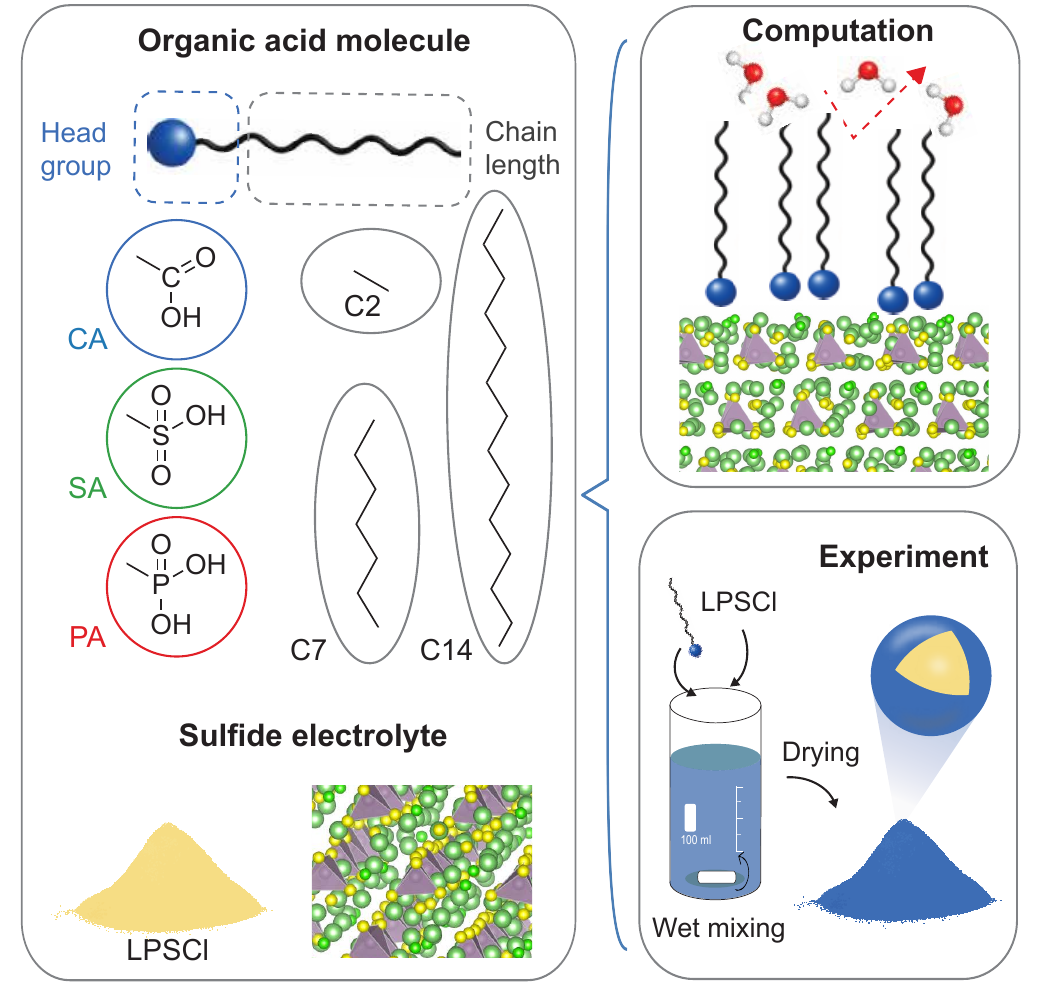}
    \caption
    {Overview of combined computational and experimental approach towards a detailed understanding of molecular surface engineering for SSE. }
    \label{fig:overview}
\end{figure}

More recently, \emph{foundation} MLIPs~\cite{mazitov_pet-mad_2025,batatia_foundation_2025,liang_nep89_2025,yuan_foundation_2026} appeared, such as PET-MAD,~\cite{mazitov_pet-mad_2025} which is built on the Point Edge Transformer architecture (PET)~\cite{pozdnyakov_smooth_2023} and trained on the chemically diverse Massive Atomic Diversity (MAD) dataset.~\cite{mazitov_massive_2025} PET-MAD proved to be sufficiently accurate to allow direct application to a variety of different systems, including SSE.~\cite{mazitov_pet-mad_2025,chorna,tisi_mechanistic_2026,fadillah_molecular_2025}
Here we fine-tune PET-MAD to LPSCl surfaces and organic coating adsorbates using an
r$^2$SCAN reference dataset (Sec.~\ref{subsec:training_dataset}). Starting from a
foundation potential pretrained across chemically diverse atomic environments provides
a transferable representation that can subsequently be adapted to the specific
LPSCl--organic interfaces considered here. This strategy enables the large-scale,
finite-temperature simulations of coated surfaces and water interfaces that underpin
this study.
In this work, we present an integrated design study of molecular coatings for SSEs, combining targeted experiments with large-scale molecular dynamics simulations to reveal how coating molecules dynamically behave and interact with water at realistic interfaces.
 Fig.~\ref{fig:overview} shows an overview of the studied systems. Based on our previous work showing promising performance of phosphonic acid (PA) head groups with long alkyl chains,~\cite{fadillah_molecular_2025} we expanded the head group selection to include also carboxylic acid (CA) and sulfonic acid (SA) to study how head group chemistry affects surface binding. Additionally, we selected alkyl chains with different lengths to investigate the effect on surface morphology. 
The molecules are denoted according to the alkyl chain length and head group, where \textit{C\textit{n}} indicates the total number of carbon atoms in the alkyl chain and the suffix specifies the head group.
Machine-learning-driven atomistic simulations establish the design rules step by step: we first identify the relevant LPSCl surface, investigate how different head groups interact with the surface, analyze how the adsorbed chains organize as a function of alkyl-chain length and surface coverage, and finally quantify how the resulting molecular layer protects the surface from water penetration and affects ionic conductivity. 
At every step, theoretical predictions are complemented with relevant experimental trends, such as conductivity measurements of the coated SSE, interfacial adhesion (FTIR and TGA), and the chain-length dependence of moisture protection (H$_2$S evolution), for carboxylic, sulfonic, and phosphonic acids of varying chain lengths.

    


\section{Methods}
\label{sec:method}

\subsection{Potential architecture and dataset generation}
\label{subsec:training_dataset}
The MLIP architecture used in this work is based on the PET architecture~\cite{pozdnyakov_smooth_2023} as implemented in metatrain.~\cite{bigi_metatensor_2026}
Unlike many commonly used MLIP frameworks,~\cite{batzner_e3-equivariant_2022,zeng_deepmd-kit_2023,musaelian_learning_2023,behler_generalized_2007} the base PET model operates directly on the Cartesian coordinates of neighboring atoms, avoiding the need to construct symmetric invariant or equivariant descriptors. This design provides a favorable compromise between model expressivity and computational cost. Rotational invariance is instead learned during training through data augmentation. Recent work~\cite{langer_probing_2024} showed that non-symmetrized PET can learn rotational symmetry and predict observables with a high degree of rotational equivariance. For further details of the architecture see the supplementary information (SI) or the relevant publications.~\cite{pozdnyakov_smooth_2023,mazitov_pet-mad_2025}

To obtain an accurate description of LPSCl surfaces and their interactions with
coating molecules, we fine-tuned a PET-MAD model on a material-specific dataset
containing 8058 structures. Fine-tuning was performed by low-rank adaptation (LoRA),~\cite{hu_lora_2022} in which the pretrained
weights are kept fixed and low-rank corrections to them are optimized.
Reference energies and forces were obtained from single-point DFT calculations using
the Vienna Ab initio Simulation Package (VASP) code~\cite{kresse_efficiency_1996}
with the renormalized Strongly Constrained and Appropriately Normed (r$^2$SCAN)
functional.~\cite{sun_scan_2015,furness_accurate_2020,ehlert_r2scan-d4_2021}
Gaussian smearing with a width of 0.08~eV was applied, together with D3-BJ dispersion
corrections~\cite{grimme_consistent_2010,Grimme2011D3BJ} and dipole corrections along the surface-normal
direction for slab calculations.

The dataset contains bulk LPSCl and LPS, clean LPSCl and LPS surfaces, isolated molecular species, and combined surface--adsorbate structures. The initial bulk \ce{Li6PS5Cl} structure was obtained from the Materials Project database.~\cite{10.1063/1.4812323} Surface slabs were generated using \texttt{pymatgen}~\cite{RN3} by cleaving low-index crystallographic orientations with Miller indices up to 3. For each surface orientation, multiple terminations were considered by shifting the cleavage position along the surface normal. During slab construction, terminations that directly break P--S bonds within the \ce{PS4^{3-}} tetrahedra were avoided. A 5~nm vacuum region was added normal to the slab to suppress interactions between periodic images.

The molecular part of the dataset includes organic and inorganic adsorbates relevant to surface functionalization and degradation. Surface-adsorbate configurations were generated by initially placing the molecules above the LPSCl surface, followed by relaxation and molecular-dynamics sampling with the PET-MAD universal potential.~\cite{mazitov_pet-mad_2025} To increase the diversity of thiophosphate surface environments, structures from previous work on lithium thiophosphate surfaces were also included and recomputed using the same DFT reference protocol.~\cite{tuerk_reconstructions_2025} 

The final dataset was split into training, validation, and test subsets using an 8:1:1 ratio. Model accuracy was assessed on the test set by comparing ML-predicted energies and forces with the corresponding DFT reference values. The fine-tuned PET-MAD model achieves an energy RMSE of 5.53~meV~atom$^{-1}$ and a force RMSE of 78.17~meV~\AA$^{-1}$. 
For this system, fine-tuning a foundation model is clearly advantageous with respect to training from scratch, leading to faster training convergence and lower errors (compare Section~\ref{sec:si_ml_validation} for a detailed analysis).

\subsection{Simulation protocol}
\label{subsec:v2_simulation_protocol}

All simulations used the fine-tuned PET-MAD potential. Molecular dynamics simulations were performed in LAMMPS~\cite{thompson_lammps_2022} with a time step of 1~fs and a Nos\'e--Hoover thermostat. The bulk LPSCl cell was first relaxed at 300~K in the NPT ensemble to obtain the equilibrium lattice parameters. Surface slabs were then constructed using these bulk-relaxed cell parameters and equilibrated at 300~K by 200~ps of NVT dynamics, followed by energy minimization. This procedure allows Li redistribution and surface reconstruction while keeping the slab cell consistent with the relaxed bulk reference used for surface-energy evaluation (definition in SI Sec.~\ref{sec:si_energies}).

Adsorption models were built by placing protonated and lithiated carboxylate, sulfonate, and phosphonate molecules of controlled chain length approximately 5~\AA{} above the relaxed LPSCl $\{111\}$ surface (468-atom slab, lateral cell of $20.3\times17.6$~\AA$^2$), in the vacuum region, with the head group oriented toward the surface. The surface--adsorbate systems were relaxed by 500~ps of NVT dynamics at 300~K followed by minimization, after which adsorption energies were evaluated (SI Sec.~\ref{sec:si_energies}). With 1, 9, or 18 adsorbed molecules, these systems contain 490-1310 atoms.

Water penetration was simulated by placing a pre-relaxed liquid water layer of 92 molecules approximately 5~\AA{} above the coated surface, followed by 1~ns of MD simulation at 300~K (750--1590 atoms in total). Water molecules whose oxygen reached an $\mathrm{O}$--Li distance below 2.5~\AA{} were counted as penetrating the coating; the resulting areal density, normalized by that of the uncoated LPSCl surface, gives the relative water accessibility, which provides a dynamic accessibility metric rather than a reaction energy.

\subsection{Surface-stability descriptor} 
To quantify how strongly the local surface coordination environment deviates from the bulk reference, we define an area-normalized coordination-deviation descriptor, $D$. This construction is motivated by previous works showing that surface stability can often be rationalized in terms of broken bonds, under-coordination, or local coordination descriptors,~\cite{Galanakis2002BrokenBond,Yasumura2023UnsaturatedCoordination,Ma2020GeometricDescriptor,CalleVallejo2023GCN} and recent work on lithium thiophosphate surfaces.~\cite{tuerk_reconstructions_2025} For LPSCl, however, a generic broken-bond count is not sufficient, because the relevant surface stability is controlled by whether chemically distinct anion environments preserve their bulk-like Li coordination. We therefore define $D$ as a bulk-referenced Li-coordination deviation around three anion environments:
\begin{equation}
    D =
    \frac{
    \sum_{\alpha \in \{\mathrm{S}_{\mathrm{PS_4}},\,\mathrm{S}_{\mathrm{iso}},\,\mathrm{Cl}\}}
    N_{\alpha}^{\mathrm{surf}}
    \left(
    1 -
    \frac{
    \overline{\mathrm{CN}}_{\alpha}^{\mathrm{surf}}
    }{
    \overline{\mathrm{CN}}_{\alpha}^{\mathrm{bulk}}
    }
    \right)
    }{
    A
    }.
    \label{eq:D_descriptor}
\end{equation}
Here, $\alpha$ denotes one of three chemically distinct anion environments: sulfur atoms belonging to $\mathrm{PS_4^{3-}}$ tetrahedra, isolated sulfide-like sulfur atoms, and chlorine atoms. $N_{\alpha}^{\mathrm{surf}}$ is the number of surface atoms of type $\alpha$, $\overline{\mathrm{CN}}_{\alpha}^{\mathrm{surf}}$ is the average Li coordination number of this surface atom type,  $\overline{\mathrm{CN}}_{\alpha}^{\mathrm{bulk}}$ is the corresponding bulk reference value for the same chemical environment, and $A$ is the surface area of the supercell. Surface atoms were defined as S and Cl atoms within $d_{\mathrm{surf}}=3.5$~\AA{} of either vacuum-facing surface.

For reference, bulk LPSCl gives a time-averaged $D$ close to zero, with a standard deviation of
$1.36\times10^{-3}~\mathrm{\AA^{-2}}$. Surfaces with positive values of $D$ indicate a larger loss of bulk-like Li coordination, while negative values indicate local Li over-coordination.

Sulfur atoms were classified according to their local P coordination. A sulfur atom was assigned to $\mathrm{S}_{\mathrm{PS_4}}$ if it belonged to a $\mathrm{PS_4^{3-}}$ tetrahedron, identified by the presence of a neighboring P atom within the P--S bonding cutoff. Sulfur atoms without such P coordination were assigned to $\mathrm{S}_{\mathrm{iso}}$. This distinction avoids mixing the Li coordination environments of framework sulfur and isolated sulfide-like sulfur, which have different bulk coordination references.
For each atom type, Li coordination numbers were calculated using a smooth distance-based cutoff, whose detailed definition is given in SI~\ref{app:coordination_number}. The descriptor $D$ should therefore be interpreted as a relative, bulk-referenced coordination-deviation density rather than a direct count of broken bonds.
A value of $D=0$ corresponds to bulk-like average Li coordination in the analyzed surface region. Positive values indicate reduced Li coordination relative to the corresponding bulk environment, while small negative values can occur when surface relaxation brings mobile Li atoms closer to S or Cl sites and increases their smooth coordination weight. By separating $\mathrm{S}_{\mathrm{PS_4}}$, $\mathrm{S}_{\mathrm{iso}}$, and Cl contributions, $D$ directly measures whether the cleaved and relaxed surface preserves the chemically distinct local coordination motifs of bulk LPSCl.

\subsection{Materials and sample preparation}
All chemicals and solvents were used without any further purification and characterization unless otherwise specified. The argyrodite $\mathrm{Li_6PS_5Cl}$ was obtained from NEI Corporation and all handling of electrolyte powder was performed in an argon glovebox ($\mathrm{O_2~\&~H_2O<0.5~ ppm}$). Propionic acid, hexanoic acid, octanoic acid, pentadecanoic acid, sodium octylsulfonate, and octylphosphonic acid were obtained from Sigma-Aldrich or Fluorochem.
Octyl sulfonic acid was obtained by using an ion-exchange resin (Dowex 50W-X8) pretreated with HCl, followed by low-temperature vacuum drying. 

Prelithiated acids were obtained by dissolving the acid in double-distilled (DD) water followed by stoichiometric addition of 1 N LiOH to neutralize each acidic proton.
 LPSCl coated with different acids was prepared by wet mixing the corresponding acids at specific molar percentages relative to LPSCl, in order to keep the ratio between the acid head group and the electrolyte surface area constant. For 500 mg of LPSCl, the respective (Li-)acid was dissolved by stirring overnight in 2 mL of dry tetrahydrofuran (THF) ($\mathrm{H_2O < 10~ppm}$, determined by Karl Fischer titration). LPSCl was then added, and the resulting suspension was stirred for 1 h to ensure uniform coating. For standard samples used in chemical characterization, the THF was subsequently evaporated overnight at 80~\textdegree C. To evaluate adhesion, the LPSCl–acid suspension was transferred to a centrifuge tube and centrifuged for 10 min at 4200 rpm, followed by the addition of 2 mL of fresh THF, which was repeated 2 more times to completely remove unreacted acid molecules.

\subsection{Experimental characterization}
Raman imaging was carried out using a WITec Alpha300R confocal Raman microscope (WITec, Ulm, Germany) equipped with Zeiss objectives (10× or 20×). The system was coupled via optical fiber to a 532 nm diode laser with a maximum output power of 125 mW. Samples were placed in airtight holders sealed with polyethylene film to prevent exposure during measurement.
Fourier transform infrared - attenuated total reflection (FTIR-ATR) spectra were recorded on a PerkinElmer Frontier spectrometer or on an ALPHA II Bruker spectrometer inside the glovebox ($\mathrm{O_2~\&~H_2O<0.5~ ppm}$) in the range of 4000-250 cm$^{-1}$. 
Thermal gravimetric analysis (TGA) was performed using Mettler Toledo TGA/DSC 3 using 40 $\mu$l aluminum crucibles between 25 \textdegree C and 600 \textdegree C at a heating ramp of 10 \textdegree C min$^{-1}$ under nitrogen gas.
The ionic conductivity of LPSCl, both with and without additives, was evaluated by electrochemical impedance spectroscopy at 25 \textdegree C using a BioLogic VSP‑300 potentiostat. Measurements were performed inside a glovebox over a frequency range of 1 Hz to 7.0 MHz with an applied AC perturbation of 10 mV. 
Approximately 30 mg of LPSCl powder was uniaxially pressed into an 8 mm diameter pellet using an ASC‑T cell (Sphere Energy) under a load of 2 tons.  Ionic conductivity was calculated as $\sigma = L/(R A)$, where $L$ is the pellet thickness (cm), $R$ is the measured resistance ($\Omega$), and $A$ is the electrode area (cm$^{2}$). The resulting conductivity in S cm$^{-1}$ was converted to mS cm$^{-1}$ by multiplying by 1000.

$\mathrm{H_2S}$ evolution was measured using a home-built setup consisting of a plastic desiccator, a modular humidity generator (MHG32-ProUmid) and a $\mathrm{H_2S}$ sensor (MODBUS RS485). After stabilization of the relative humidity (RH) at 20\% the $\mathrm{H_2S}$ evolution was measured.

\section{Results}

\subsection{LPSCl surface stability}
\label{subsec:surface_energy_result}

\begin{figure*}[t]
    \centering
    \includegraphics[width=\textwidth]{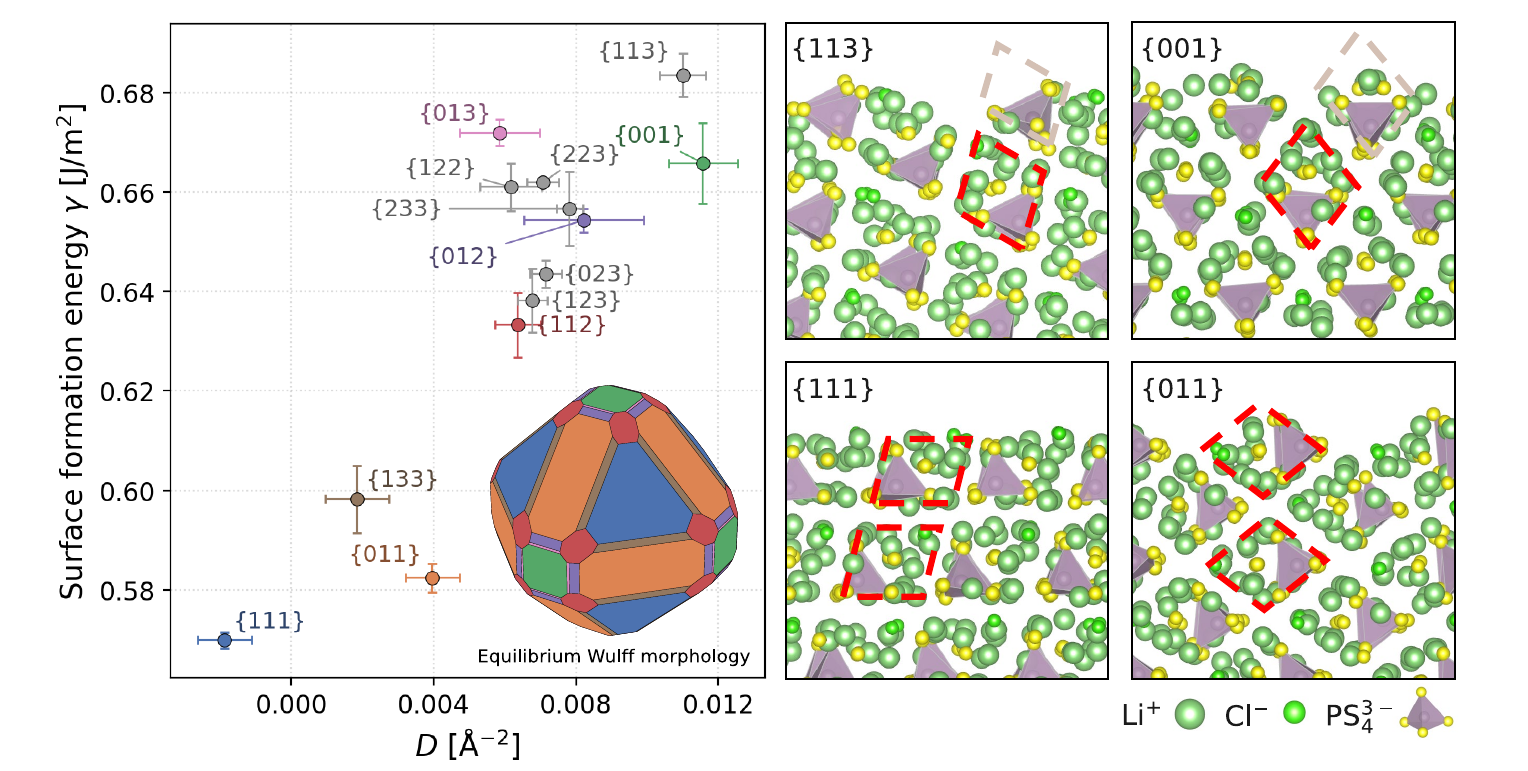}
    \caption{
Correlation between surface formation energy $\gamma$ and the signed coordination descriptor $D$ for generated LPSCl surface terminations, together with the corresponding equilibrium Wulff morphology constructed from the lowest-energy termination of each surface family. The facet colors in the Wulff construction correspond to the colors used for the highlighted surface families in the scatter plot. For each surface family, the plotted point corresponds to the lowest-energy termination among the different cleavage positions generated for that orientation. The low-energy $\{111\}$ and $\{011\}$ surfaces preserve LPSCl-like surface clusters with approximately correct \ce{Li6PS5Cl} stoichiometry. In these clusters, neighboring \ce{PS4^{3-}} tetrahedra act collectively as a structural framework that supports the Li/S/Cl environment. Red dashed boxes highlight representative preserved LPSCl-like clusters. Gray boxes highlight disrupted cluster regions in high-energy surfaces: in the $\{113\}$ surface, the cluster is cut by the surface termination, whereas in the $\{001\}$ surface, the initially exposed cluster becomes unstable and reconstructs during finite-temperature relaxation.
}
    \label{fig:D_surface_energy}
\end{figure*}

We first analyzed the relative stability of clean LPSCl surfaces generated from different crystallographic orientations using our fine-tuned PET model. For each Miller family, several terminations were constructed by shifting the cleavage position along the corresponding surface normal; Fig.~\ref{fig:D_surface_energy} shows the lowest-energy termination obtained for each orientation. The calculated surface formation energies range from approximately 0.57 to 0.68~J~m$^{-2}$. Vertical and horizontal error bars denote the standard deviations of $\gamma$ and $D$ within the selected lowest-energy termination class, with both quantities averaged over 20 minimization frames.

Among the screened surfaces, the $\{111\}$ termination exhibits the lowest surface formation energy. The corresponding Wulff construction predicts that the equilibrium morphology is dominated by the $\{011\}$ and $\{111\}$ facet families, which together account for approximately two thirds of the total surface area. This is qualitatively consistent with previous first-principles Wulff constructions of LPSCl, which also identified the symmetry-equivalent $\{110\}$/$\{011\}$ family as a dominant exposed surface.~\cite{DAmore2022Wulff} These low-energy surfaces show small values of the signed coordination descriptor $D$, indicating that the S--Li and Cl--Li coordination environments remain close to the bulk reference after relaxation (slightly negative values of $D$, as for $\{111\}$, reflect local Li compensation allowed by the smooth coordination-number definition; see SI~\ref{app:coordination_number}). The structural snapshots in Fig.~\ref{fig:D_surface_energy} provide a microscopic interpretation of this trend: surfaces with low $\gamma$ preserve LPSCl-like surface clusters with approximately correct \ce{Li6PS5Cl} stoichiometry (red dashed boxes), i.e., local structural units in which neighboring \ce{PS4^{3-}} tetrahedra form a framework supporting the Li, S, and Cl atoms. When the surface cleavage preserves these clusters without strongly breaking or twisting them, the local S/Cl coordination remains close to the bulk reference and the surface formation energy is low.

In contrast, high-energy surfaces correspond to disrupted local cluster environments (gray boxes in Fig.~\ref{fig:D_surface_energy}), through two distinct mechanisms: the $\{113\}$ termination cuts directly through an LPSCl-like cluster, whereas the $\{001\}$ surface fails to preserve its surface cluster during finite-temperature relaxation, as mobile Li atoms redistribute near the vacuum interface. Both mechanisms lead to a larger coordination deviation and a higher surface formation energy; a detailed discussion, including the role of finite-temperature Li mobility compared to static 0~K calculations,~\cite{tuerk_reconstructions_2025} is provided in SI~\ref{app:surface_stability_detail}. Because the $\{111\}$ surface has the lowest surface formation energy, retains a bulk-like local LPSCl cluster environment, and belongs to one of the two dominant facet families in the predicted equilibrium morphology, it is selected as the representative stable LPSCl surface for the subsequent adsorption-energy calculations and water-penetration simulations.

\subsection{Head group chemistry and interfacial binding}
\label{subsec:v2_binding}

\begin{figure*}[!htbp]
    \centering
    \includegraphics[width=\linewidth]{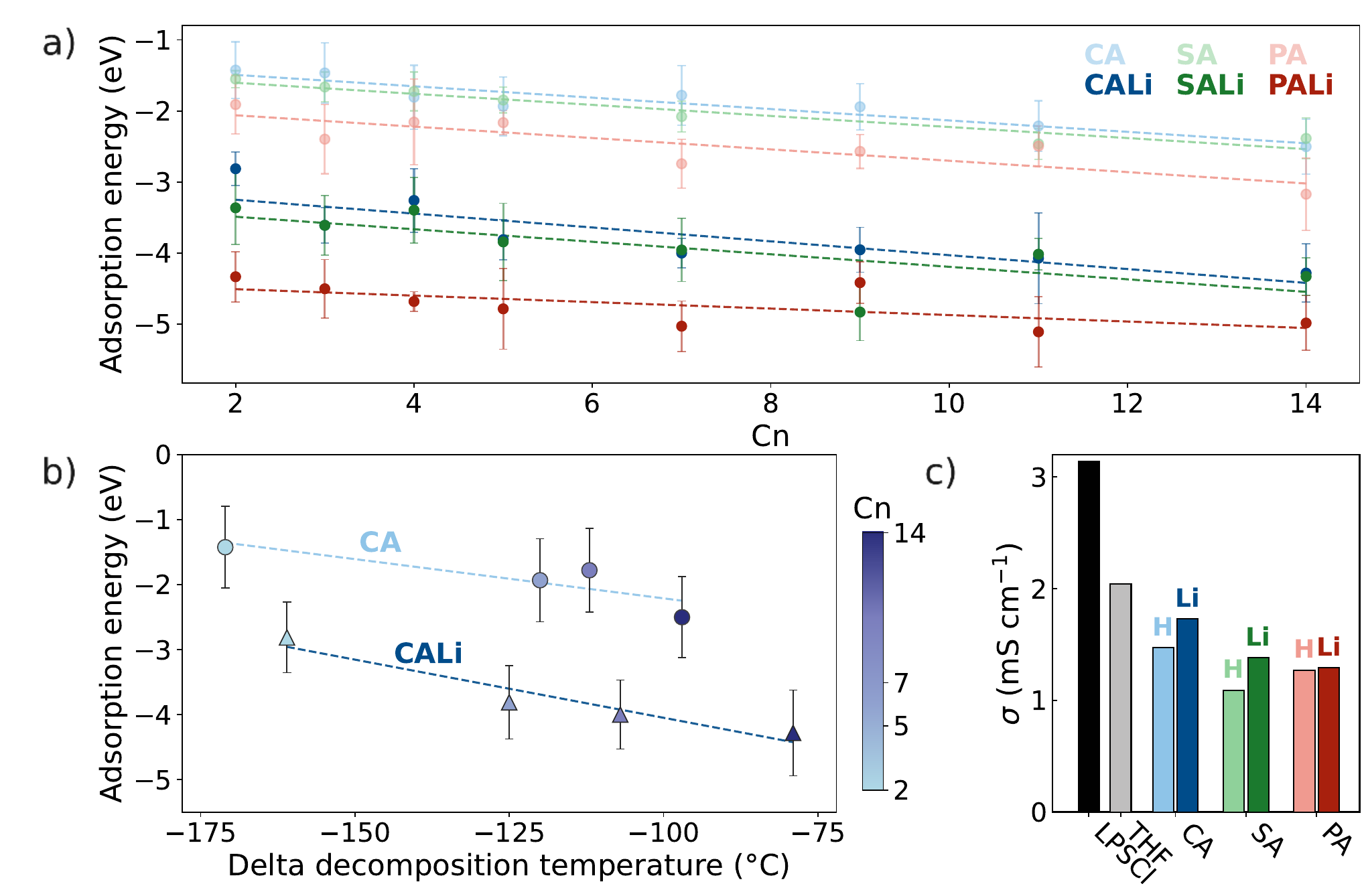}
    \caption{(a) Computed single-molecule adsorption energy versus alkyl-chain length for
    carboxylic, sulfonic, and phosphonic acids (CA, SA, PA) and their lithiated forms
    (CALi, SALi, PALi). (b) Calculated adsorption energy compared with the measured change
    in decomposition temperature for the CA/CALi series. This comparison is used as a
    qualitative indicator rather than a direct measurement of adhesion. Marker color
    encodes chain length; circles and triangles denote protonated and lithiated forms,
    respectively. (c) Experimental ionic conductivity of LPSCl before and after THF
    treatment and after coating with different head groups and lithiation states. In (a) and (b), vertical error bars denote the standard deviation of the adsorption energy over the sampled minimization frames for each condition.}
    \label{fig:v2_adsorption}
\end{figure*}

We examine the binding of the coating molecules to the LPSCl surface, focusing on the
combined effects of alkyl-chain length, head group chemistry, and lithiation state.
Fig.~\ref{fig:v2_adsorption}a shows the computed single-molecule adsorption energy as a
function of alkyl-chain length for the carboxylic, sulfonic, and phosphonic acids (CA, SA,
PA) and their lithiated forms (CALi, SALi, PALi). For all head groups, adsorption is favorable and increases with increasing chain length. This indicates that the alkyl chain also provides additional stabilization of the molecules through contact with the LPSCl surface. 

Experimentally, coating was achieved using a wet approach similar to our earlier
work (see Sec.~\ref{sec:method}).~\cite{fadillah_molecular_2025} The FTIR spectra
(Fig.~\ref{SI Fig: FTIR-XA}) show that molecular species remain associated with LPSCl
after processing. Samples prepared from protonated and prelithiated additives display
similar head-group regions after coating, which is compatible with a change in protonation
state during processing, but does not by itself establish complete in-situ lithiation.
Raman spectra (Fig.~\ref{SI Fig: Raman}) indicate that the main LPSCl vibrational features
are retained after THF treatment and coating.

For the carboxylate series, Fig.~\ref{fig:v2_adsorption}b shows a qualitative association
between the calculated adsorption energy and the change in decomposition temperature
relative to the corresponding additive alone. More strongly adsorbing carboxylates tend to
show smaller decreases in decomposition temperature after association with LPSCl. However,
the number of experimental points is limited, and the decomposition temperature also
depends on the intrinsic thermal properties of the free and surface-associated species.
The TGA comparison should therefore be regarded as qualitative support rather than as a
direct experimental measurement of adsorption energy. The head-group-dependent TGA data
are provided in Fig.~\ref{SI Fig: T_decomp}, and qualitatively corroborate the
computed adsorption-energy ordering among CA, SA, and PA.

Within the tested surface models, the lithiated species generally show more favorable adsorption energies than their protonated counterparts. A possible explanation is that the Li atoms associated with the lithiated head groups can participate in coordination with both the head-group O atoms and nearby surface S/Cl sites, whereas the acidic H atoms mainly form more localized contacts with surface anions (SI  Section~\ref{sec:si_head_rdf}). Differences among the head groups may similarly reflect differences in their electronic structure, bond polarity, acidity, charge distribution, and oxygen-donor ability, in addition to the number and arrangement of available O coordination sites and the local adsorption geometries sampled at the surface.~\cite{Raman2010OrganicFilm,Albrechts2025TitaniaAdsorption}
 In particular, PA/PALi tends to show the most negative adsorption energies, while the differences between CA and SA are smaller. These trends should nevertheless be interpreted cautiously because the head groups can adopt different configurations and protonation states, and the available experiments do not isolate intrinsic adsorption sufficiently to establish a definitive head-group ranking. The calculations are therefore used to identify broad trends rather than a universal ordering of binding strength. Additional head-H/head-Li RDFs and representative adsorption configurations are provided in the Supporting Information (Figs.~\ref{fig:v2_head_bulk_rdf} and~\ref{fig:si_adsorption_snapshots}) as qualitative structural context.

The finite-temperature simulations also directly show spontaneous deprotonation events for the sulfonic acid head group, with the detached H relocating toward sulfide-like surface S sites. This observation is consistent with the comparatively high proton lability of the SA head group and provides a possible microscopic pathway for proton transfer during surface binding. Together with the similarity of the coated FTIR spectra obtained from protonated and prelithiated additives, it is compatible with deprotonation during coating or solvent processing.~\cite{Braks2024-TCA,fadillah_molecular_2025} The fact that deprotonation occurs spontaneously over the course of short simulations provides strong evidence for a strong thermodynamic driving force, even though the statistics are not sufficient to determine the actual deprotonation rate. The analysis in SI Section~\ref{sec:si_deprotonation} is presented as direct evidence of a deprotonation event within the simulated SA trajectories%
.

The ionic conductivity after coating with different head groups, protonated and lithiated, was measured experimentally and is shown in Fig.~\ref{fig:v2_adsorption}c.  The decrease in conductivity after THF treatment indicates that the solvent-processing step itself affects LPSCl, potentially through mild chemical interactions and/or solvent-induced changes in particle morphology that increase grain-boundary resistance.~\cite{ruhl_impact_2021,Kim2021-cosolvent,Lee2017-solvent}
Prelithiated additives generally retain
higher conductivity than the corresponding protonated additives, while the conductivity
trend does not follow the computed adsorption energies in a one-to-one manner. This lack of
direct correspondence may reflect several coupled effects, including interfacial chemistry,
amount of surface-bound organic molecules, particle contacts, and local Li mobility. The simulations provide a nanoscale picture of how head-group chemistry and lithiation state modify the interfacial environment underlying the experimentally observed differences in conductivity retention.
For further analysis of the role of chain length and surface coverage, this study concentrates on CALi, as it provides the most favorable practical balance between calculated adsorption and conductivity retention among the tested
head groups, rather than because it is the strongest-binding species.

\subsection{Chain-length-dependent coating morphology}
\label{subsec:v2_prep}

\begin{figure}[!htbp]
    \centering
    \includegraphics[width=\linewidth]{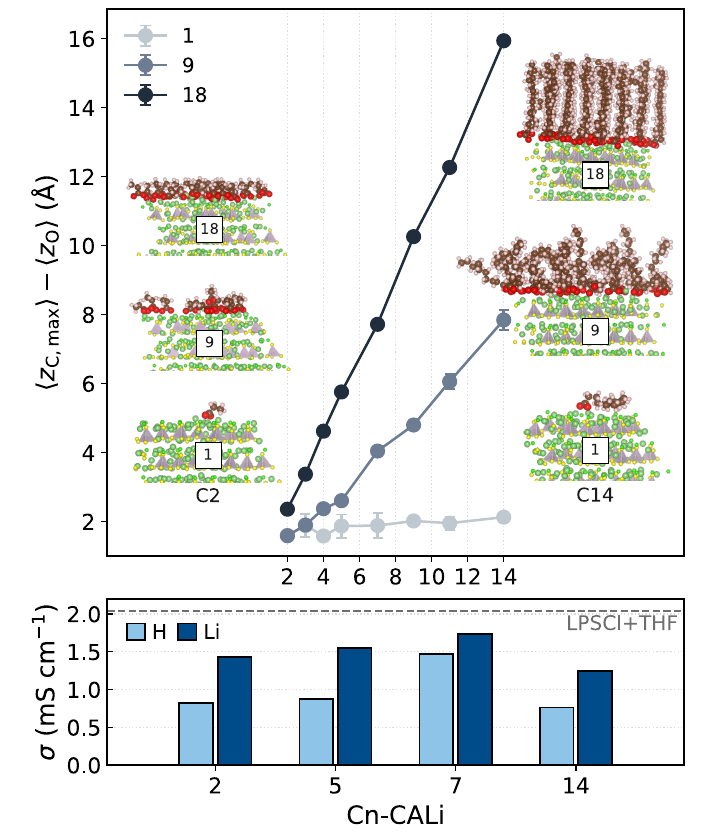}
    \caption{Coverage- and chain-length-dependent orientation of lithiated carboxylate
    chains on the LPSCl $\{111\}$ surface, together with the corresponding experimental
    conductivity trend for Cn-CALi coatings. The vertical molecular extension, quantified
    as $\langle z_{\mathrm{C,max}}\rangle - \langle z_{\mathrm{O}}\rangle$, is shown for
    Cn-CALi molecules at coverages of 1, 9, and 18 adsorbed molecules. Representative MD
    snapshots illustrate the transition from flat-lying isolated molecules at low coverage
    to upright, brush-like layers for long chains at high coverage. The bottom panel shows
    the ionic conductivity of Cn-CALi-coated LPSCl as a function of chain length. The dashed horizontal line indicates the conductivity of the THF-treated LPSCl reference.}
    \label{fig:ZCn}
\end{figure}

Building on the adsorption-energy analysis and the head group comparison above, we next
examine how the adsorbed molecular layer organizes as a function of alkyl-chain length and
surface coverage. We use lithiated carboxylates on the LPSCl $\{111\}$ surface as a
representative model system.

Fig.~\ref{fig:ZCn} examines how the conformation of carboxylate chains changes with
alkyl-chain length and coverage for lithiated carboxylate. The vertical molecular extension
is quantified by $\langle z_{\mathrm{C,max}}\rangle - \langle z_{\mathrm{O}}\rangle$, where
$z_{\mathrm{C,max}}$ is the height of the highest carbon atom in the alkyl chain, and
$z_{\mathrm{O}}$ is the height of the anchoring oxygen atoms. For a single adsorbed
$\mathrm{CH_3(CH_2)}_{n-1}\mathrm{COOLi}$ molecule, this height remains small and changes
only weakly with chain length. This trend is illustrated by the bottom snapshots in
Fig.~\ref{fig:ZCn}: both the short-chain (C2, bottom left) and long-chain (C14,
bottom right) molecules lie nearly flat on the surface after 500~ps relaxation. Thus,
increasing the chain length does not make an isolated molecule stand up; instead, the
longer chain remains in contact with the substrate, consistent with the chain-length
dependence of the adsorption energy discussed in Section~\ref{subsec:v2_binding}.

The behavior changes as surface coverage increases. For nine adsorbed molecules, the
molecular height increases with chain length, indicating that intermolecular packing begins
to lift the alkyl chains away from the surface. The middle snapshots in Fig.~\ref{fig:ZCn}
show this transition directly: the short-chain molecules (C2, middle left) form a
compact layer close to the surface, whereas the long-chain molecules (C14, middle right)
form a thicker and more disordered adsorbed layer. For eighteen adsorbed molecules, the
vertical extension increases much more strongly with chain length. This is consistent with
the top snapshots, where the short-chain system (C2, top left) remains relatively thin,
while the long-chain system (C14, top right) forms an upright, dense molecular layer.
Therefore, the molecular orientation is controlled by both chain length and surface
coverage. At low coverage, chain-surface contact favors flat-lying configurations, whereas
at high coverage, chain-chain packing drives the chains into more upright conformations.

The effects of this coverage-dependent structure can be seen in the lower panel of Fig.~\ref{fig:ZCn}, which shows the experimental ionic
conductivity of Cn-CALi-coated LPSCl as a function of alkyl-chain length. A denser or thicker
organic layer may improve surface coverage but can also introduce additional resistance to
Li-ion transport across grain boundaries. The dependence
on alkyl chain length is relatively weak but systematic: short-chain additives lead to
stronger conductivity losses, likely due to stronger interaction with, or penetration into,
the sulfide framework, as supported by FTIR and conductivity analysis
(Fig.~\ref{SI Fig: FTIR-Cx-CA} and Fig.~\ref{SI Fig: EIS-C7-CA}). In contrast, very long chains (e.g., C14) also reduce
conductivity, which is attributed to hindered Li$^+$ transport through the non-polar alkyl
region and the higher effective mass loading at constant molar ratios.
Among the four chain lengths tested (C2, C5, C7, and C14), ionic transport is best preserved for C7.

\subsection{Moisture protection: water penetration and H$_2$S evolution}
\label{subsec:v2_moisture}

\begin{figure*}[!ht]
    \centering
    \includegraphics[width=\linewidth]{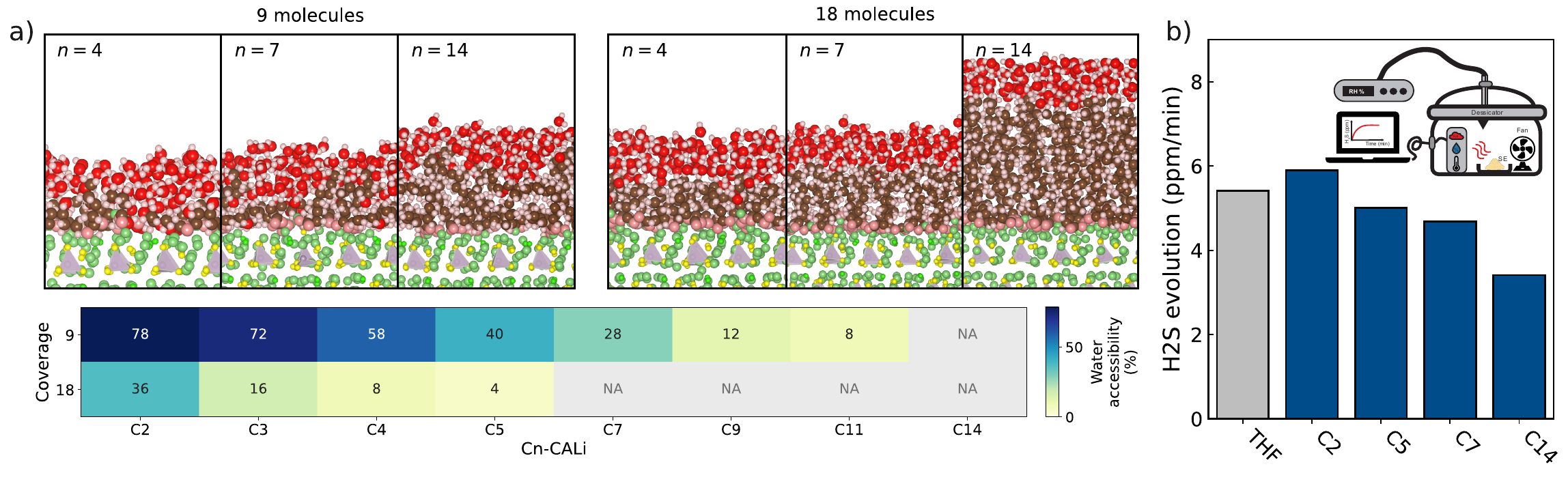}
    \caption{Moisture protection from simulation and experiment. 
(a) Simulated water penetration through lithiated carboxylate coatings on LPSCl $\{111\}$:
MD snapshots for $n=4,7,14$ at coverages of 9 and 18 molecules, and a heatmap of water
accessibility versus chain length and coverage. Water accessibility is defined as the
average number of water molecules forming short-range $\mathrm{O_{water}}$--Li contacts
($<2.5$~\AA), normalized by the corresponding value for the uncoated surface. Water and
carboxylate oxygen atoms are shown in dark and light red, respectively. 
(b) Initial H$_2$S evolution rate extracted from time-resolved H$_2$S measurements for
10~mol\% coated LPSCl at 20\% RH, together with a THF-treated uncoated
reference.}
    \label{fig:v2_moisture}
\end{figure*}

Finally, we assess whether the coated layers protect the surface from moisture by combining the atomistic water-penetration simulations with the experimental degradation rate in a
single comparison (Fig.~\ref{fig:v2_moisture}).

Computationally (Fig.~\ref{fig:v2_moisture}a), the coatings are compared by the relative water accessibility defined in Sec.~\ref{subsec:v2_simulation_protocol}. Within the lithiated-carboxylate model,
water protection depends strongly on chain length and surface coverage: once the head group
is anchored, the morphology of the hydrocarbon layer becomes decisive for suppressing water
access. At intermediate coverage, short and medium chains still allow water to reach the
surface, whereas longer chains progressively reduce water accessibility by filling the
interfacial gaps. At high coverage, the same suppression is achieved at shorter chain
lengths. Notably, irrespective of surface coverage, the onset of the water protection in both cases (C14 and C7 at coverages of 9 and 18 molecules, respectively, see Fig.~\ref{fig:v2_moisture}) is at a coating thickness of approximately 8~\AA{} (compare Fig.~\ref{fig:ZCn} and Fig.~\ref{fig:si_height_water}), provided that the chain length is sufficient to reach this thickness. Since coating thickness is jointly determined by chain length and surface coverage, these two parameters act as coupled descriptors of the protective layer.
The simulations, therefore, identify chain length and surface coverage as the two variables controlling the continuity of the hydrocarbon barrier and the resulting accessibility of water to reactive surface sites.

Experimentally, the consequences of water penetration are reflected in the amount of H$_2$S released into a closed chamber.
Fig.~\ref{fig:v2_moisture}b shows the initial H$_2$S evolution rates for
10~mol\% coated LPSCl at 20\% RH, together with a THF-treated uncoated
reference,
extracted from the time traces in
Fig.~\ref{fig:si_h2s_evolution}. Short-chain coatings
remain close to this level, whereas longer chains progressively lower the
initial H$_2$S slope, in qualitative agreement with the simulated suppression
of water penetration. The experimental reduction is less pronounced than the idealized simulations
would suggest, as expected for powder-processed coatings that are likely to be
inhomogeneous and may not reach the high-density, defect-free coverage represented by the
highest-coverage simulation models.
Moisture protection thus improves with increasing chain length, with C14 giving the lowest initial H$_2$S evolution rate, whereas ionic conductivity is best preserved for C7 (Fig.~\ref{fig:ZCn}). Since protection sets in only once the coating reaches the thickness identified above and thicker layers increasingly hinder Li-ion transport, the most favorable balance among the tested chain lengths is expected for the shortest chains that reach this thickness, i.e. in the intermediate regime between C5 and C14.

\section{Conclusions}
\label{sec:conclusion}

In this work, we presented a framework that combined large-scale MLIP simulation with targeted experiments to establish molecular design principles for moisture-protective coatings on SSEs. 
For the Li$_6$PS$_5$Cl system, we identified the low-energy $\{111\}$ surface as a representative interface for analyzing adsorption, coating morphology, conductivity retention, and water accessibility. We further demonstrate the utility of foundation MLIPs as priors for modeling complex atomistic systems, showing that fine-tuned models converge more rapidly and achieve lower energy errors than models trained from scratch on the same dataset.

The calculations show that interfacial anchoring depends on head-group chemistry and lithiation state, with lithiated species generally exhibiting more favorable adsorption than their protonated counterparts. Differences among the head groups are comparatively smaller and should be interpreted as broad trends rather than a universal binding-strength ranking.

Once the head group is anchored, the organization of the alkyl chains becomes critical for moisture protection. Isolated molecules remain largely flat on the surface, whereas increasing coverage drives intermolecular packing and progressively more upright, brush-like configurations. Longer chains and higher coverage reduce direct water access to the Li-rich surface, and the same level of protection can be achieved with shorter chains as the coating density increases. The experimental H$_2$S measurements follow the same chain-length trend, supporting the water-exclusion mechanism revealed by the simulations.

Importantly, moisture protection and ionic transport do not vary in the same way with chain length.
While the water barrier improves progressively with increasing chain length and becomes significant at coating thicknesses of approximately 8~\AA{}, the measured ionic conductivity shows a non-monotonic dependence with a maximum around C7.
This 
suggests that intermediate chain lengths offer the best balance between moisture exclusion and Li-ion transport.
This optimum corresponds to the shortest chains that still reach the protective thickness, while longer chains form thicker layers that increasingly hinder Li-ion transport.

Taken together, these results show that effective molecular coatings require a balance between interfacial anchoring, hydrophobic surface coverage, and preservation of Li-ion transport. The simulations provide critical insights for experimental trends that are otherwise probed only indirectly, and establish chain length and surface coverage as coupled design variables, acting through the coating thickness, for controlling the protection--transport trade-off.
This study combines simulations and experiments to identify trends and mechanistic insights into a complex passivation process, and validates the use of fine-tuned universal MLIPs to enable atomistic simulations of unprecedented scale and complexity at first-principles accuracy.
The weak dependence of H$_2$S evolution on coating suggests that future studies should focus on understanding the mechanisms of both water and Li transport across the coating, as well as the impact of defects and partial coverage. The availability of a validated MLIP provides the necessary infrastructure to facilitate these challenging joint modeling--experimental investigations, which have the potential to enable rational selection of passivation layers tailored to specific
electrolyte compositions and processing conditions, accelerating
the path toward commercially viable moisture-tolerant
SSEs.

\section{Data Availability}
The data to reproduce the results in the paper, such as input files and the model used to perform the simulations, will be available on Materials Cloud~\cite{talirz_materials_2020} upon publication.

\begin{acknowledgments}
A.C., M.C., D.T., and L.B. acknowledge support from the Swiss National Science Foundation (SNSF) (Grant No. Sinergia CRSII5\_202296). H.T. acknowledges funding by the Deutsche Forschungsgemeinschaft (DFG, German Research Foundation) as a Walter-Benjamin Fellow (Project No. 534232519). L.B. thanks Prof. Salentinig from the University of Fribourg for the use of the Confocal Raman Spectrophotometer.
This work was supported by grants from the Swiss National Supercomputing Centre (CSCS) under Project No. lp95.

\end{acknowledgments}

\bibliography{others,referencesDT}


\onecolumngrid
\newpage

\begin{center}
  {\large\bfseries Supplementary Material for\par}
  \vspace{6pt}
  {\large\bfseries Combining Simulations and Spectroscopy to Design Protective Coatings on Sulfide Solid Electrolytes\\}
  \vspace{12pt}
{Zhiyi Wang, Leonie Braks, Michele Ceriotti, Ali Coskun, Hanna T\"{u}rk, and Davide Tisi}
\end{center}
\vspace{12pt}

\setcounter{page}{1}

\setcounter{section}{0}
\setcounter{figure}{0}
\setcounter{table}{0}
\setcounter{equation}{0}
\setcounter{page}{1}

\renewcommand{\thesection}{S\arabic{section}}
\renewcommand{\thesubsection}{\thesection.\arabic{subsection}}
\renewcommand{\thesubsubsection}{\thesubsection.\arabic{subsubsection}}
\makeatletter
\def\p@subsection{}
\def\p@subsubsection{}
\makeatother
\renewcommand{\thefigure}{S\arabic{figure}}
\renewcommand{\thetable}{S\arabic{table}}
\renewcommand{\theequation}{S\arabic{equation}}

\renewcommand{\theHsection}{SI.\arabic{section}}
\renewcommand{\theHfigure}{SI.\arabic{figure}}
\renewcommand{\theHtable}{SI.\arabic{table}}
\renewcommand{\theHequation}{SI.\arabic{equation}}

\section{Computational model validation and surface analysis}

\subsection{Machine-learning potential training and validation}
\label{sec:si_ml_validation}

The calculations were performed using machine-learning interatomic potentials based
on the Point Edge Transformer (PET) architecture.~\cite{pozdnyakov_smooth_2023}
PET has demonstrated high accuracy across molecular and materials
benchmarks.~\cite{pozdnyakov_smooth_2023,mazitov_pet-mad_2025,tisi_mechanistic_2026}

The model used in this work was initialized from the pretrained universal PET-MAD v1.0.1 checkpoint~\cite{mazitov_pet-mad_2025} and subsequently fine-tuned on the LPS/LPSCl dataset described in
Sec.~\ref{subsec:training_dataset}. The network has two message-passing layers, each containing a two-layer transformer with
an embedding dimension of 256 and 8 attention heads, and uses a cutoff radius of 4.5~\AA.
Fine-tuning was performed by low-rank adaptation (LoRA)~\cite{hu_lora_2022} with rank 8: the pretrained
weights were kept fixed and only the low-rank updates (16\,384 parameters, 0.42\% of the
model) were optimized. Training used batches of 850 atoms; the model used for all simulations is the checkpoint at epoch 1100, selected on the validation set as described below. Energy and force mean-squared errors entered the loss
with weights of 0.1 and 1.0, respectively.

\begin{figure*}[!htbp]
    \centering
    \includegraphics[width=\linewidth]{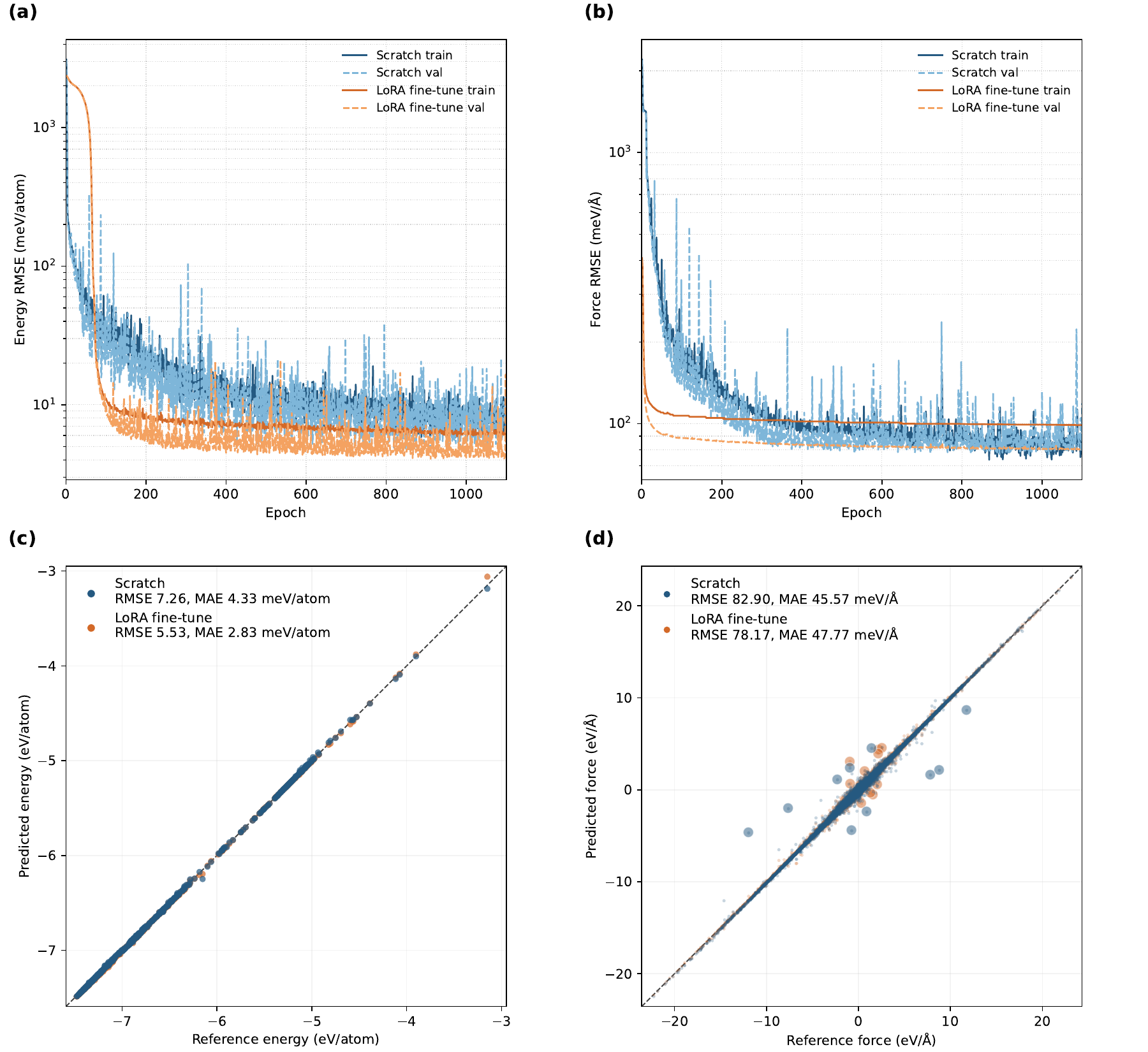}
    \caption{
    Comparison of the PET architecture trained from random initialization
    (Scratch) and by LoRA fine-tuning from the pretrained universal PET-MAD
    v1.0.1 checkpoint (LoRA fine-tune), trained with identical settings. (a) Energy and (b) force RMSE on the training (solid) and validation (dashed)
    sets as a function of epoch. (c) Energy and (d) Cartesian force-component parity plots on the held-out test set
    for the checkpoints selected on the validation set (epoch 900 for Scratch and epoch 1100 for LoRA fine-tune); the ten largest force errors of each model are shown as
    larger markers, and dashed lines indicate perfect agreement.
    }
    \label{fig:scratch_finetune}
\end{figure*}

To assess the benefit of pretrained initialization, the same PET architecture was
also trained from random initialization using the same material-specific dataset
and evaluated on the same held-out test set. For this comparison, the model trained from scratch and the
LoRA fine-tuned model were trained with identical data and settings (learning-rate schedule, batch of 850 atoms,
cutoff, loss weights and random seed); the only difference is the initialization. Both models were trained for more than 1400 epochs,
with a checkpoint saved every 50 epochs. For each training approach, the reported model is not the last checkpoint
but the one with the lowest validation loss among all checkpoints up to epoch 1100 (epoch 900 for the model
trained from scratch and epoch 1100 for the fine-tuned model).
Note also that beyond 1100 epochs the model trained from scratch
starts to overfit: its training error keeps decreasing faster than the validation error. On the other hand, the fine-tuned
model is converged at this point: the validation energy and force RMSEs of all later checkpoints up to
epoch 1500 differ from those of the selected checkpoint by less than 0.4~meV~atom$^{-1}$ and
0.6~meV~\AA$^{-1}$, respectively. As shown in
Fig.~\ref{fig:scratch_finetune}, 
fine-tuning reduces the energy RMSE
from 7.26 to 5.53~meV~atom$^{-1}$ (about 25\%) and the corresponding MAE from 4.33 to 2.83~meV~atom$^{-1}$. The force errors of the two models are of similar magnitude, with RMSEs of 82.90~meV~\AA$^{-1}$ (scratch) and 78.17~meV~\AA$^{-1}$ (fine-tuned) and
MAEs of 45.6 and 47.8~meV~\AA$^{-1}$, respectively. The small difference in force RMSE comes mainly from a few
large errors (enlarged markers in Fig.~\ref{fig:scratch_finetune}d): the largest force-component errors of the
model trained from scratch reach 7.4~eV~\AA$^{-1}$, mostly on O atoms in a single test structure: an off-stoichiometric Li--O--P slab containing O--O bonded
species, with DFT forces of up to 15~eV~\AA$^{-1}$ on these atoms. This chemistry is represented by only 15 of
the 6448 training structures, none with the same composition, and the fine-tuned model predicts all atoms of
this structure with errors below 1.4~eV~\AA$^{-1}$. Excluding the 20 atoms with the largest errors, the
force RMSEs of the two models are 74.4 (scratch) and 76.1~meV~\AA$^{-1}$ (fine-tuned). The fine-tuned model also
converges faster and more smoothly: after 100 epochs its validation
energy RMSE is 9.6~meV~atom$^{-1}$, compared with 33.5~meV~atom$^{-1}$ for the model trained from scratch, which
needs about 500 epochs to reach a similar value (11.0~meV~atom$^{-1}$; values averaged over 20 epochs).
Overall, fine-tuning lowers the energy error, gives fewer extreme force outliers and converges much faster,
while typical force errors remain of similar magnitude. This suggests that the representations acquired
during pretraining on the broad MAD dataset~\cite{mazitov_massive_2025} provide a useful prior for local
environments that are only sparsely represented in the material-specific dataset, rather than requiring all
relevant local interactions to be learned from scratch.

\subsection{Coordination-number definition}
\label{app:coordination_number}

Coordination numbers were computed using a smooth distance-based switching function. For an atom $i$, the coordination number with respect to neighboring atoms $j$ is
\begin{equation}
    \mathrm{CN}_i = \sum_j s(r_{ij}),
    \label{eq:appendix_CN}
\end{equation}
where $r_{ij}$ is the interatomic distance and
\begin{equation}
s(r)=
\begin{cases}
1, & r \leq d_0, \\[4pt]
\dfrac{1}{2}
\left[
\cos\left(
\pi \dfrac{r-d_0}{r_0}
\right)
+1
\right],
& d_0 < r < d_0+r_0, \\[8pt]
0, & r \geq d_0+r_0 .
\end{cases}
\label{eq:appendix_switch}
\end{equation}
In this work,
\begin{equation}
    d_0=0.5~\text{\AA},
    \qquad
    r_0=3.0~\text{\AA}.
\end{equation}
Atoms below 0.5~\AA{} therefore contribute with weight 1, atoms between 0.5 and 3.5~\AA{} contribute with a smoothly decreasing weight, and atoms beyond 3.5~\AA{} are not counted.

The species-resolved coordination numbers used in Eq.~\ref{eq:D_descriptor} are
\begin{equation}
    \mathrm{CN}_{i}^{\mathrm{S-Li}}
    =
    \sum_{j\in \mathrm{Li}} s(r_{ij}),
    \qquad
    i\in \mathrm{S},
\end{equation}
and
\begin{equation}
    \mathrm{CN}_{i}^{\mathrm{Cl-Li}}
    =
    \sum_{j\in \mathrm{Li}} s(r_{ij}),
    \qquad
    i\in \mathrm{Cl}.
\end{equation}
The same switching function was used for bulk and slab structures, so that
\begin{equation}
    \frac{
    \overline{\mathrm{CN}}_{\alpha}^{\mathrm{surf}}
    }{
    \overline{\mathrm{CN}}_{\alpha}^{\mathrm{bulk}}
    }
\end{equation}
measures the surface coordination relative to the bulk reference for species $\alpha$.

\subsection{Interpretation of the surface descriptor}
\label{app:surface_stability_detail}

For some terminations, including $\{111\}$, the signed coordination descriptor $D$ becomes slightly negative. This does not imply a negative number of broken bonds. Because the coordination number is smooth, Li atoms moving closer to surface S or Cl sites during relaxation increase their fractional coordination weight. A small negative $D$ therefore indicates local Li compensation or slight over-coordination, and the relative trend across terminations is more meaningful than the absolute value.

The present protocol combines 300~K bulk equilibration, 300~K NVT slab relaxation, and final minimization. It therefore captures Li redistribution and surface reconstruction that are suppressed in static 0~K calculations. This interpretation is consistent with previous work showing that molecular-dynamics-induced reconstruction, surface density, and local coordination affect the stability of lithium thiophosphate surfaces.~\cite{tuerk_reconstructions_2025}

\subsection{Surface termination selection}
\label{sec:si_termination_selection}

\begin{figure*}[t]
    \centering
    \includegraphics[width=\linewidth]{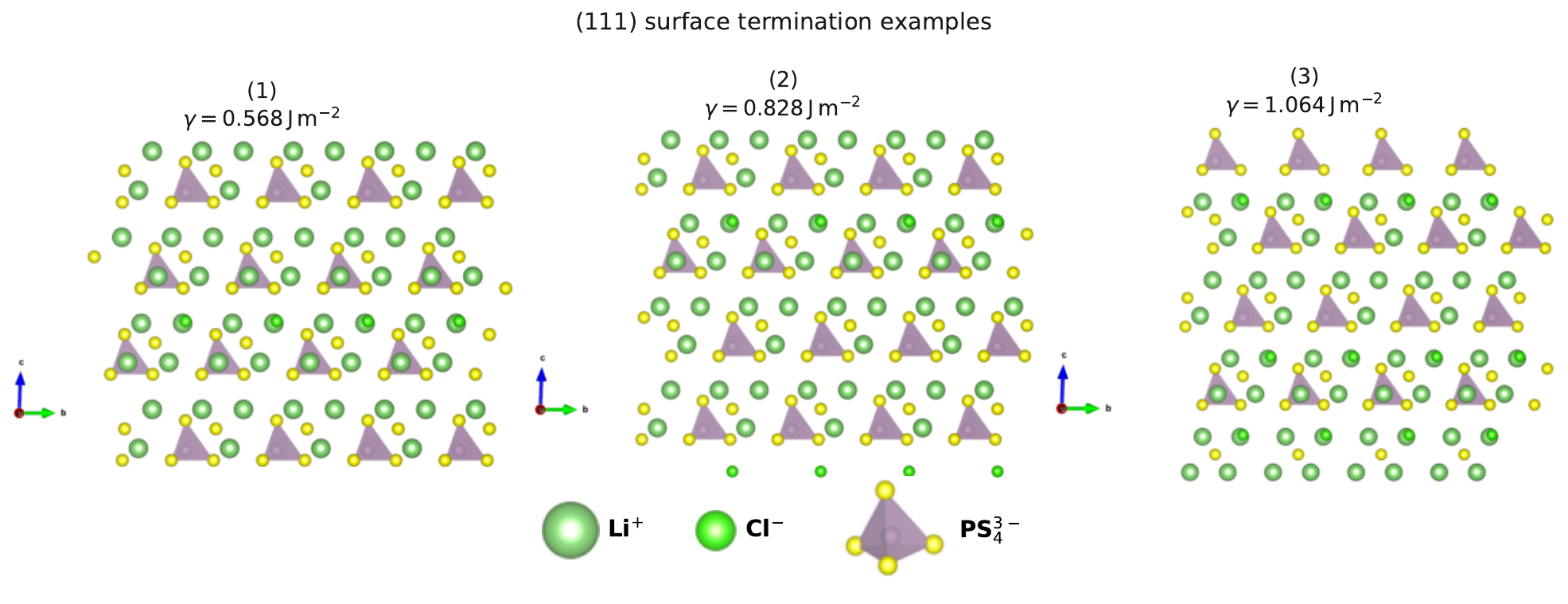}
    \caption{
    Representative $\{111\}$ surface terminations generated by shifting the cleavage
    position along the surface normal. The different terminations exhibit substantially
    different surface formation energies because of changes in the local surface
    stoichiometry and coordination environment. The lowest-energy termination is used as
    the representative $\{111\}$ surface in the surface-stability analysis and Wulff
    construction.
    }
    \label{fig:si_111_termination_examples}
\end{figure*}

For a given Miller family, different cleavage positions generate distinct surface
terminations with different local stoichiometries and coordination environments.
Fig.~\ref{fig:si_111_termination_examples} illustrates this for the $\{111\}$ family,
where representative terminations span a wide range of surface formation energies,
from 0.568 to 1.064~J~m$^{-2}$.

Because the equilibrium morphology is determined by the most stable surface available for
each orientation, we use the lowest-energy termination of each Miller family in the Wulff
construction and in the comparison of surface stabilities. Higher-energy terminations are
therefore treated as metastable cleavage configurations rather than representative
equilibrium facets.

\subsection{Robustness of the Wulff morphology}
\label{sec:si_wulff_convergence}

To assess the sensitivity of the predicted equilibrium morphology to the set of
surface orientations included in the Wulff construction, we progressively expanded
the input from the three $\{001\}$, $\{011\}$, and $\{111\}$ families to all
low-index surfaces with $\max(h,k,l)\leq2$, and finally to the complete set of
surface families considered in this work.

\begin{figure*}[t]
    \centering
    \includegraphics[width=\linewidth]{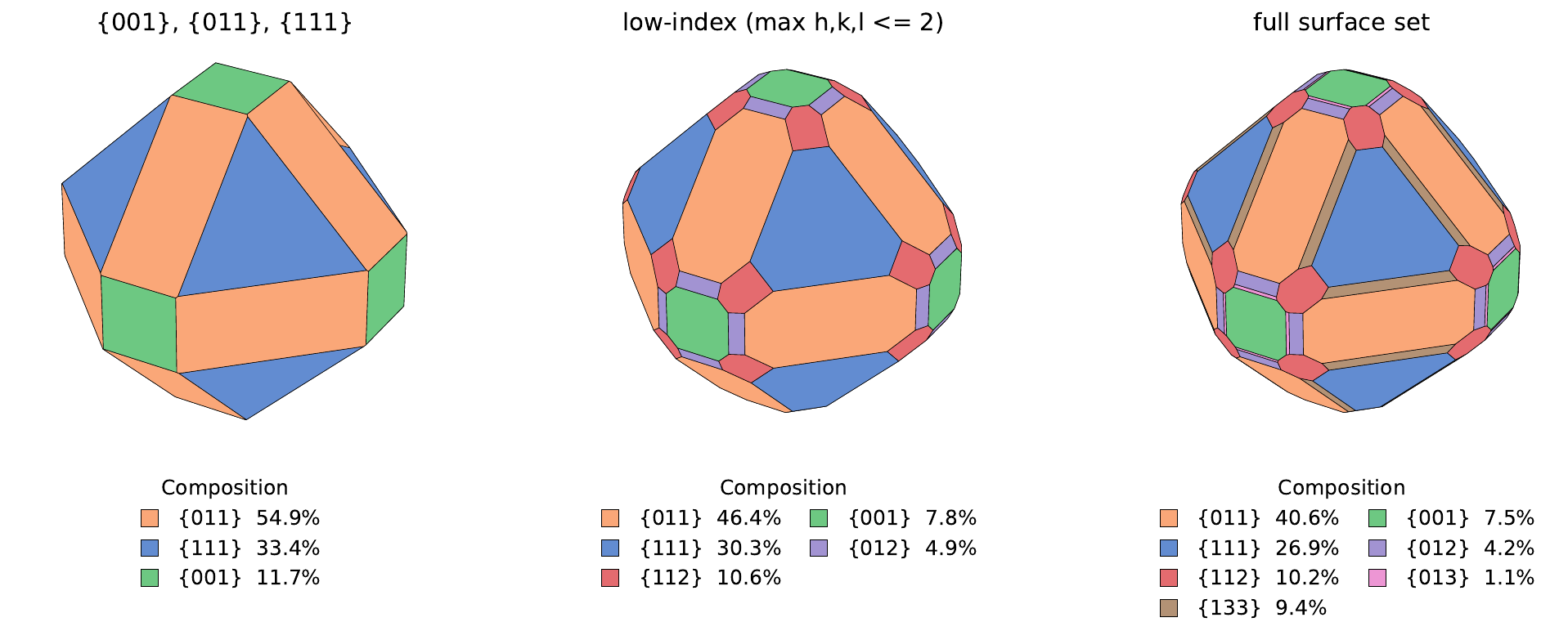}
    \caption{
    Sensitivity of the predicted LPSCl Wulff morphology to the set of surface
    orientations included in the construction. From left to right, the Wulff shape
    is constructed using only $\{001\}$, $\{011\}$, and $\{111\}$; all low-index
    surfaces with $\max(h,k,l)\leq2$; and the complete surface set considered in
    this work. Additional higher-index facets progressively subdivide the
    equilibrium morphology, while the $\{011\}$ and $\{111\}$ families remain the
    two dominant surface contributions.
    }
    \label{fig:si_wulff_sets}
\end{figure*}

As shown in Fig.~\ref{fig:si_wulff_sets}, additional surface orientations
progressively subdivide the equilibrium shape and reduce the individual area
fractions of the dominant facets. Nevertheless, the qualitative morphology remains
unchanged: the $\{011\}$ and $\{111\}$ families are the two largest contributors in
all three constructions. In the full surface set, they account for 40.6\% and
26.9\% of the total surface area, respectively, corresponding to approximately
two thirds of the predicted equilibrium surface.

\subsection{Surface and adsorption energies}
\label{sec:si_energies}

The surface formation energy of a slab with $N_{\mathrm{slab}}$ atoms and surface area $A$ (two equivalent surfaces) is
\begin{equation}
    \gamma = \frac{E_{\mathrm{slab}} - N_{\mathrm{slab}}\,E_{\mathrm{bulk}}/N_{\mathrm{bulk}}}{2A},
\end{equation}
where $E_{\mathrm{bulk}}/N_{\mathrm{bulk}}$ is the energy per atom of bulk LPSCl. The adsorption energy of $N$ molecules is
\begin{equation}
    E_{\mathrm{ads}} = \frac{E_{\mathrm{slab}+N\mathrm{mol}} - E_{\mathrm{slab}} - N\,E_{\mathrm{mol}}}{N},
\end{equation}
where $E_{\mathrm{mol}}$ is the energy of the isolated molecule; negative values indicate favorable adsorption. All energies
are averages over minimized MD frames.

\section{Head-group chemistry and interfacial binding}
\label{sec:si_experimental}

\subsection{Adsorption trends on low-energy LPSCl surfaces}
\label{sec:si_adsorption_011}

\begin{figure}[!htbp]
    \centering
    \includegraphics[width=\linewidth]{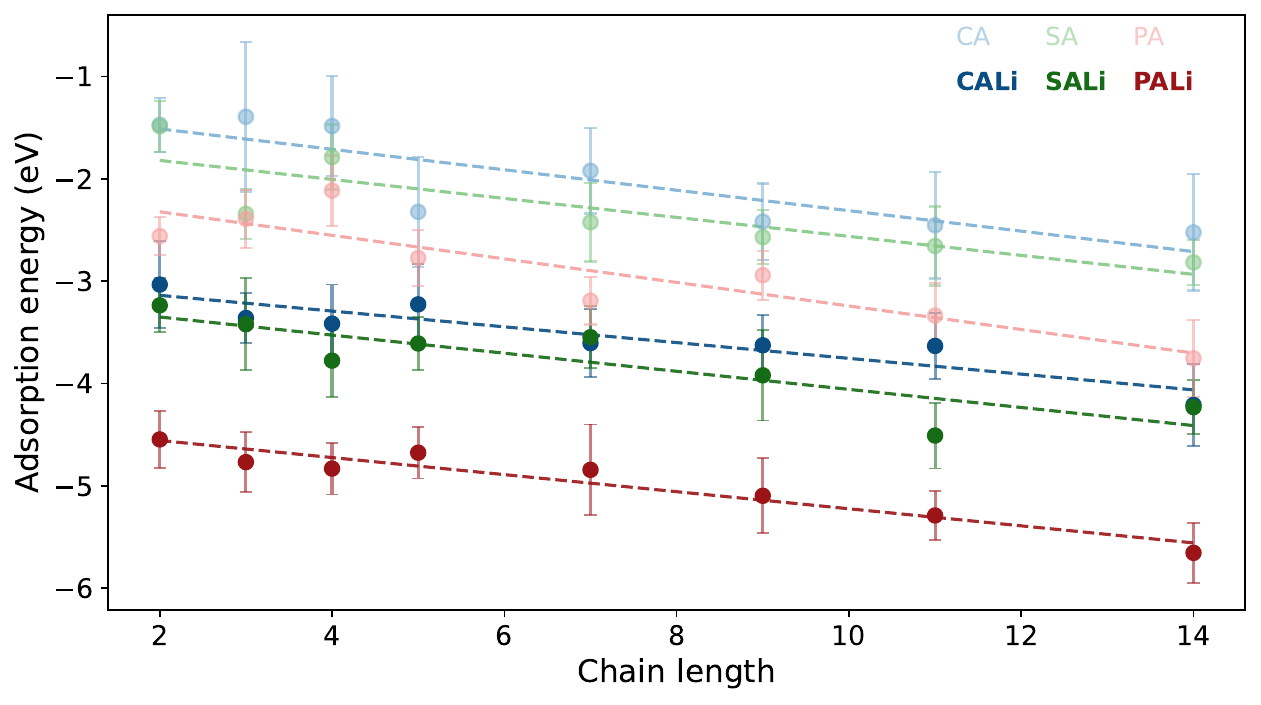}
    \caption{Adsorption energies of molecules with different head groups and alkyl-chain lengths on the LPSCl $\{011\}$ surface. The broad trends are similar to those obtained on $\{111\}$: adsorption generally becomes more favorable with increasing chain length, and the lithiated models tend to bind more strongly than their protonated counterparts. The quantitative differences among head-group families refer to the tested configurations and should not be interpreted as a universal ranking.}
    \label{fig:si_adsorption_011}
\end{figure}

To test whether the adsorption trends obtained for LPSCl $\{111\}$ are surface-specific, adsorption energies were also evaluated on the low-energy $\{011\}$ surface. Fig.~\ref{fig:si_adsorption_011} reproduces the broad tendencies observed on $\{111\}$: adsorption generally becomes more favorable with alkyl-chain length, and the lithiated models tend to adsorb more strongly than the corresponding protonated models. Phosphonate-based species give the most negative values within the tested configurations. Because the head groups can adopt different geometries and protonation states, the quantitative separation between head-group families should be interpreted cautiously.

\begin{figure}[!htbp]
    \centering
    \includegraphics[width=\linewidth]{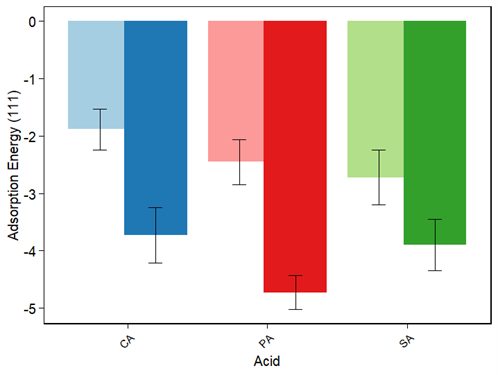}
    \caption{Average adsorption energies of C2--C14 protonated and prelithiated acids on the LPSCl $\{111\}$ surface.}
    \label{SI Fig: AdsEn_average}
\end{figure}

For comparison across the full chain-length range, Fig.~\ref{SI Fig: AdsEn_average} summarizes the average adsorption energies of the protonated and prelithiated acids on LPSCl $\{111\}$.

\subsection{Spectroscopic evidence for coating and phase stability}

\begin{figure*}[!htbp]
    \centering
    \includegraphics[width=\linewidth]{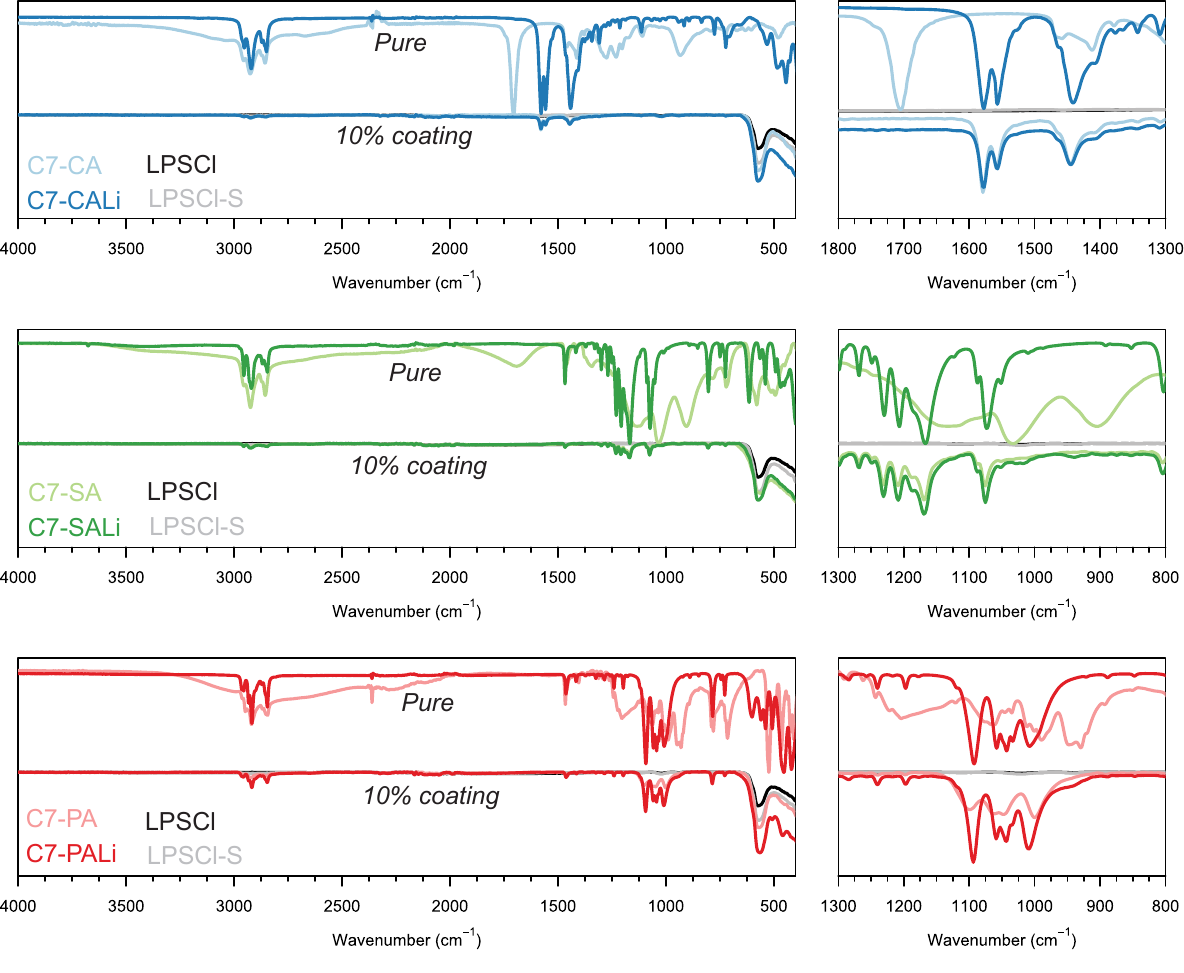}
    \caption{ATR--FTIR spectra of pristine additives and LPSCl coated with 10~mol\% of (a) C7-CA and C7-CALi, (b) C7-SA and C7-SALi, and (c) C7-PA and C7-PALi.}
    \label{SI Fig: FTIR-XA}
\end{figure*}

The ATR--FTIR spectra in Fig.~\ref{SI Fig: FTIR-XA} provide evidence that molecular species remain associated with LPSCl after processing. Samples prepared from protonated and prelithiated additives show similar head-group regions after coating, while several features of the pristine protonated acids are suppressed or shifted. These observations are compatible with changes in protonation state during processing, but the spectra alone do not establish complete in-situ lithiation or quantify the fraction of deprotonated molecules.

\begin{figure}[!htbp]
    \centering
    \includegraphics[width=0.48\linewidth]{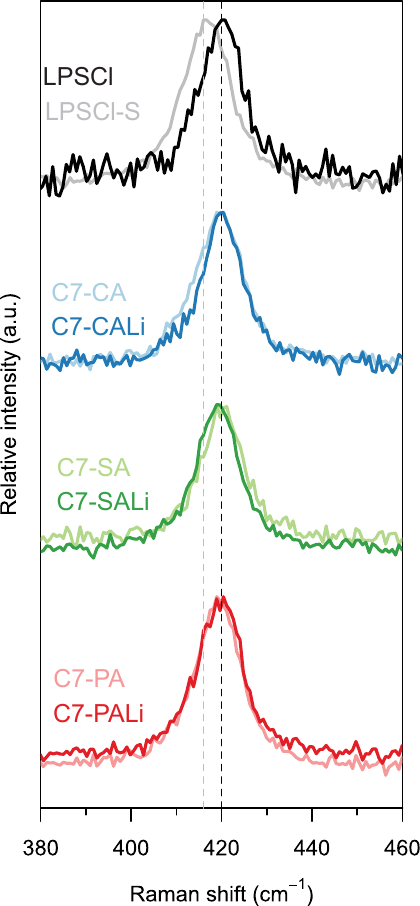}
    \caption{Raman spectra of pristine, solvent-treated, and 10~mol\%-coated LPSCl samples.}
    \label{SI Fig: Raman}
\end{figure}

Raman spectra show that the main vibrational features of LPSCl are retained after THF treatment and molecular coating (Fig.~\ref{SI Fig: Raman}).

\subsection{Thermal stability and washing-based adhesion assessment}

\begin{figure*}[!htbp]
    \centering
    \includegraphics[width=\linewidth]{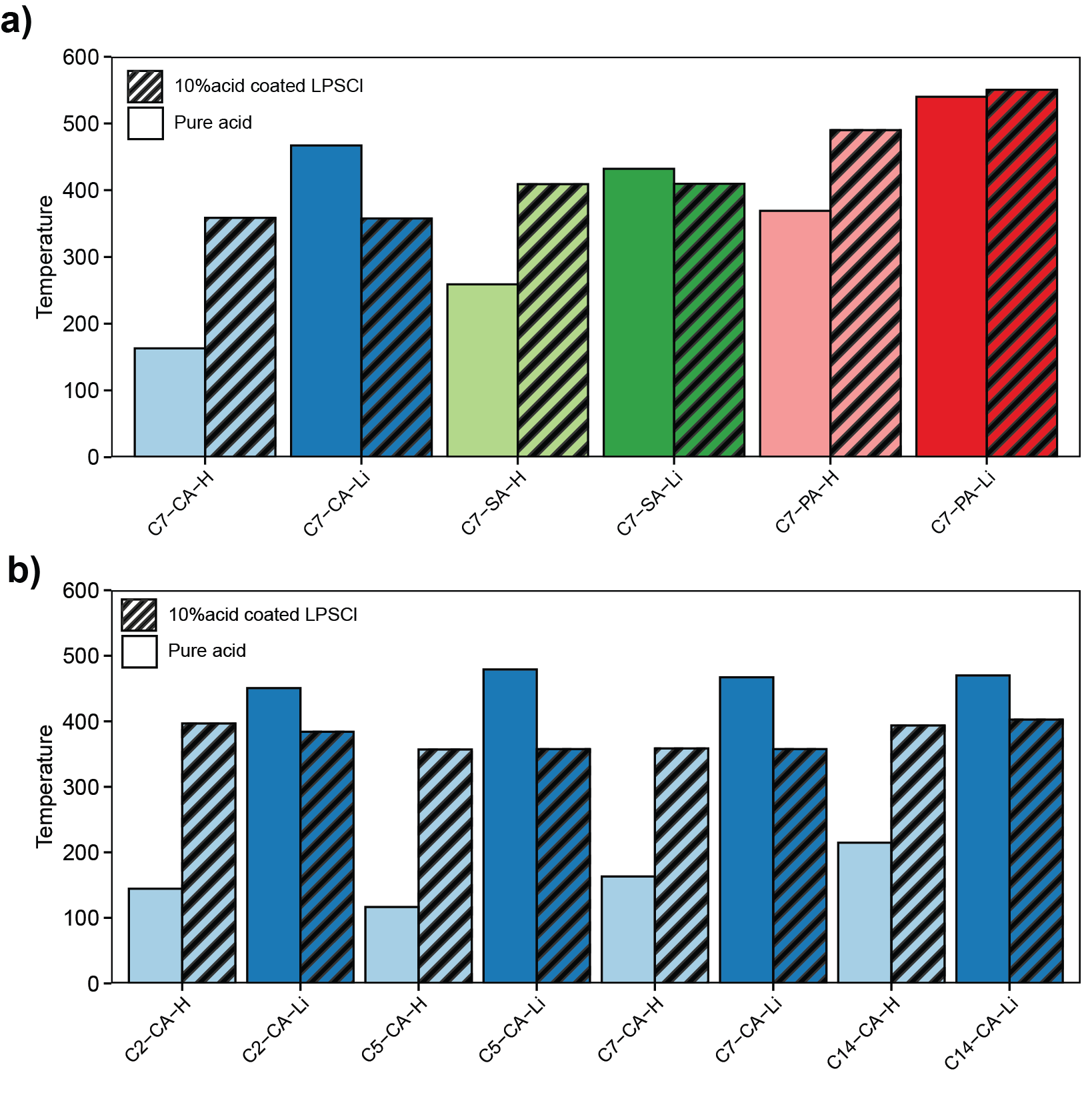}
    \caption{TGA decomposition temperatures of pristine protonated and prelithiated additives and surface-coated samples for (a) different C7 head groups and (b) carboxylic acids of different chain lengths.}
    \label{SI Fig: T_decomp}
\end{figure*}

The TGA decomposition temperatures provide an experimental comparison of the thermal stability of the free and surface-associated additives. Fig.~\ref{SI Fig: T_decomp} compares the different C7 head groups and the carboxylic-acid chain-length series. Prelithiated additives, particularly PALi, show high decomposition temperatures; however, these values also reflect the intrinsic thermal stability of the corresponding salts and therefore provide only qualitative support for the calculated ordering of adsorption energies.

\begin{figure*}[!htbp]
    \centering
    \includegraphics[width=\linewidth]{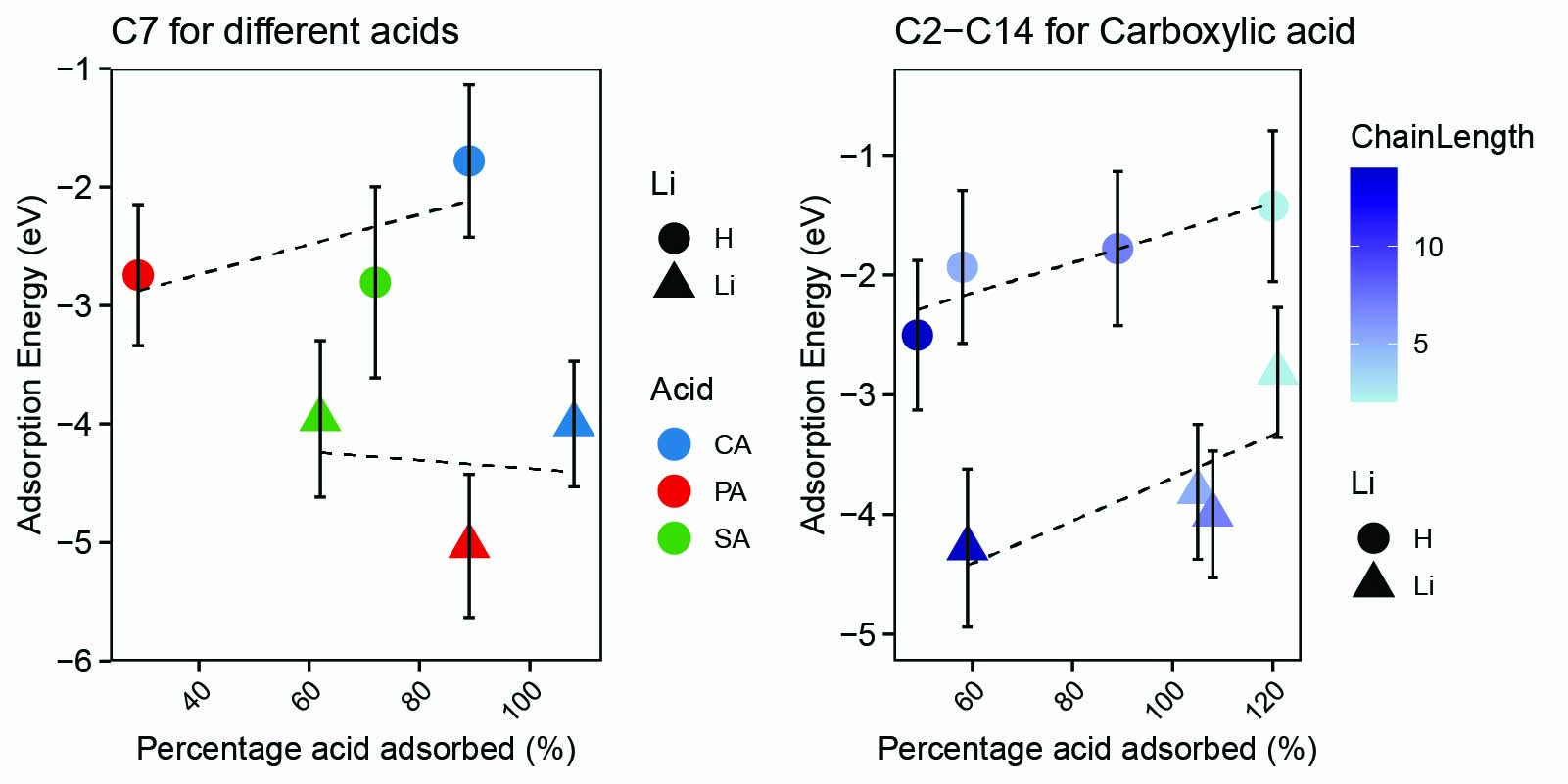}
    \caption{Correlation between the computed adsorption energies and the percentage of acid remaining after THF washing of the coated LPSCl powders.}
    \label{SI Fig: Corr_Comp_percent}
\end{figure*}

Adhesion was also assessed indirectly by washing coated particles repeatedly with THF and quantifying the remaining organic content by TGA. The correlation with the computed adsorption energies is weak, particularly for short-chain acids (Fig.~\ref{SI Fig: Corr_Comp_percent}), because the procedure can reflect kinetic removal and surface reactions rather than equilibrium adsorption. The washing data should therefore be treated as qualitative support rather than a direct adhesion measurement.

\begin{figure*}[!htbp]
    \centering
    \includegraphics[width=\textwidth]{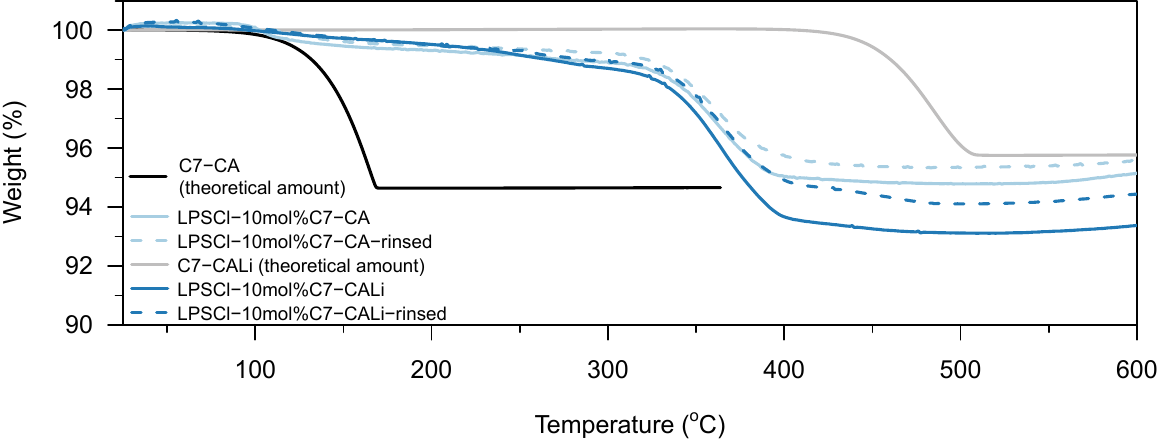}
    \caption{Thermogravimetric analysis between 25 and 600~\textdegree C for C7-CA and C7-CALi and 10~mol\%-coated LPSCl samples before and after THF washing and centrifugation.}
    \label{SI Fig:C7-CA}
\end{figure*}

The raw C7 TGA traces below show the washing response for the three head-group families used in the main comparison. Fig.~\ref{SI Fig:C7-CA} presents the carboxylic-acid series.

\begin{figure*}[!htbp]
    \centering
    \includegraphics[width=\textwidth]{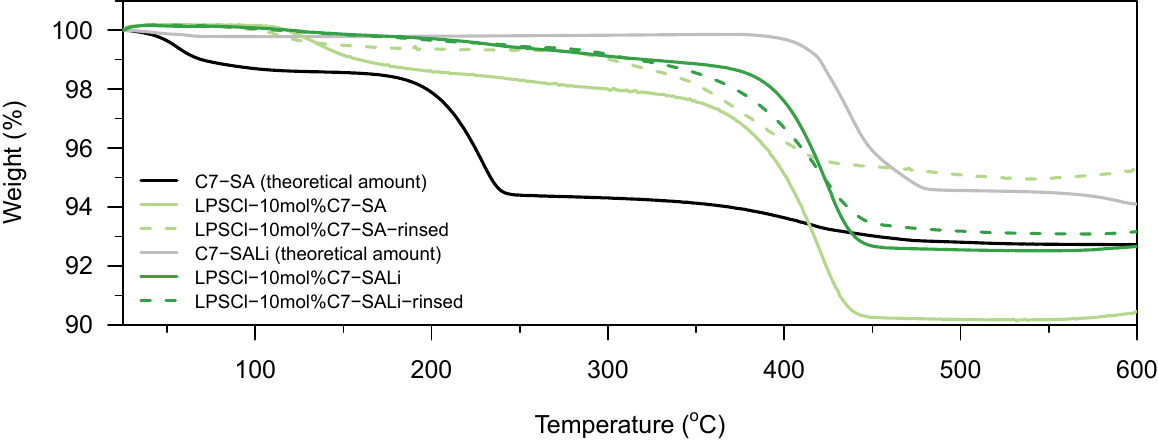}
    \caption{Thermogravimetric analysis between 25 and 600~\textdegree C for C7-SA and C7-SALi and 10~mol\%-coated LPSCl samples before and after THF washing and centrifugation.}
    \label{SI Fig:C7-SA}
\end{figure*}

The corresponding sulfonic-acid and sulfonate traces are shown in Fig.~\ref{SI Fig:C7-SA}.

\begin{figure*}[!htbp]
    \centering
    \includegraphics[width=\textwidth]{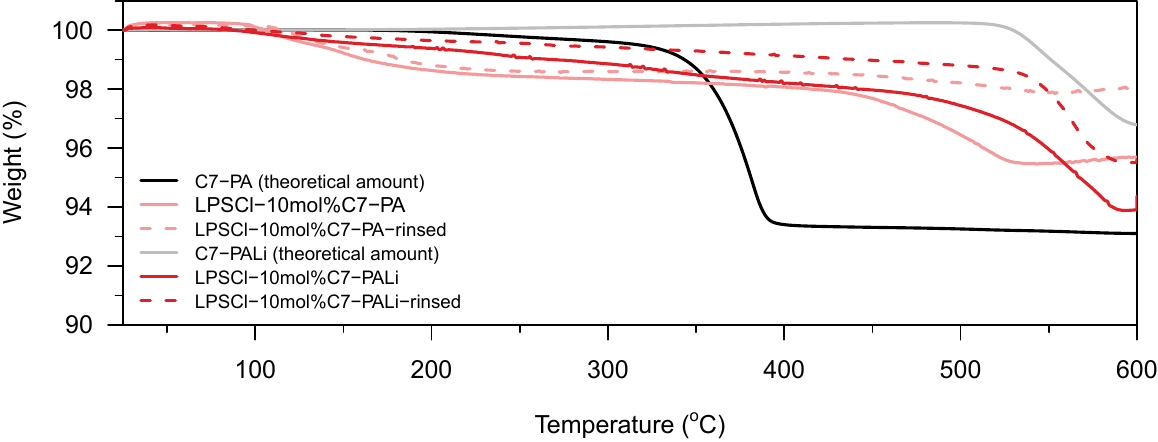}
    \caption{Thermogravimetric analysis between 25 and 600~\textdegree C for C7-PA and C7-PALi and 10~mol\%-coated LPSCl samples before and after THF washing and centrifugation.}
    \label{SI Fig:C7-PA}
\end{figure*}

Fig.~\ref{SI Fig:C7-PA} provides the equivalent phosphonic-acid and phosphonate data.

\subsection{Representative adsorption configurations}
\label{sec:si_adsorption_snapshots}

\begin{figure*}[!htbp]
    \centering
    \includegraphics[width=\linewidth]{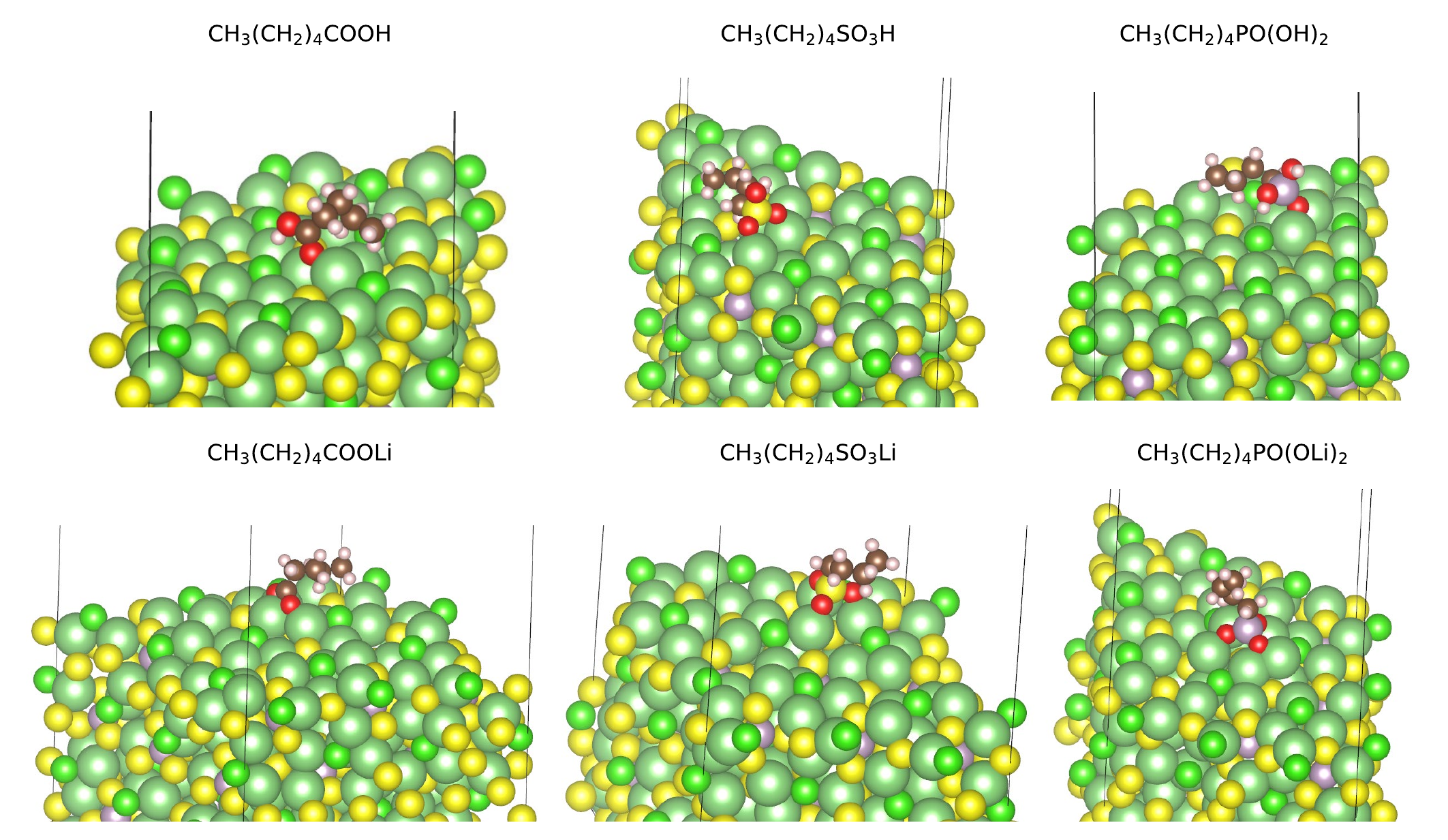}
    \caption{Representative relaxed adsorption snapshots for molecules with different head groups on the LPSCl $\{111\}$ surface. The top row shows $\mathrm{CH_3(CH_2)_4COOH}$, $\mathrm{CH_3(CH_2)_4SO_3H}$, and $\mathrm{CH_3(CH_2)_4PO(OH)_2}$. The bottom row shows $\mathrm{CH_3(CH_2)_4COOLi}$, $\mathrm{CH_3(CH_2)_4SO_3Li}$, and $\mathrm{CH_3(CH_2)_4PO(OLi)_2}$. These snapshots provide representative local configurations and are not intended to define a unique binding motif for each head group.}
    \label{fig:si_adsorption_snapshots}
\end{figure*}

Fig.~\ref{fig:si_adsorption_snapshots} shows representative relaxed structures for protonated and lithiated head groups on LPSCl $\{111\}$. The snapshots illustrate the range of local environments sampled by the different head groups, but individual configurations should not be interpreted as unique adsorption motifs or as direct evidence for the origin of the adsorption-energy differences.

\subsection{Head-H and head-Li surface-contact RDFs}
\label{sec:si_head_rdf}

\begin{figure*}[!htbp]
    \centering
    \includegraphics[width=\linewidth]{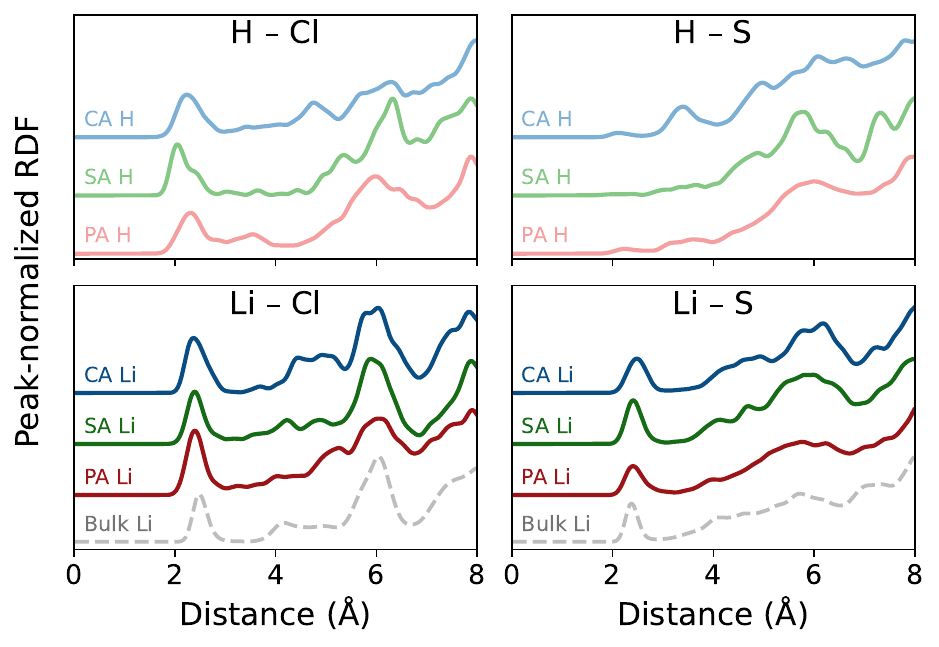}
    \caption{Peak-normalized RDFs comparing surface contacts involving head-H and head-Li
    at the LPSCl $\{111\}$ surface. The upper panels show head-H--Cl and head-H--S
    distributions for protonated CA, SA, and PA. The lower panels compare head-Li--Cl/S
    distributions in CALi, SALi, and PALi with the corresponding bulk Li--Cl/S RDFs of
    LPSCl. The RDFs provide qualitative information about sampled local environments and
    should not be interpreted as direct evidence of a unique binding mechanism or a
    quantitative explanation of adsorption-energy differences.}
    \label{fig:v2_head_bulk_rdf}
\end{figure*}

To provide additional structural context for the protonated and prelithiated models,
Fig.~\ref{fig:v2_head_bulk_rdf} compares peak-normalized RDFs involving H or Li initially
associated with the molecular head group. The protonated systems show short-range H--Cl
features and weaker or broader H--S features, whereas the prelithiated systems show
head-Li--Cl and head-Li--S distances that overlap approximately with the corresponding
bulk Li--Cl and Li--S distributions.

These RDFs indicate that protonated and prelithiated heads sample different local
interfacial environments, but they do not by themselves establish specific bonding motifs
or explain the calculated adsorption-energy differences. The curves are peak-normalized,
represent ensemble averages over several adsorption configurations, and do not include H
atoms after they detach from the molecular head group. The comparison is therefore used
only as qualitative structural information.

\subsection{Finite-time deprotonation in the SA system}
\label{sec:si_deprotonation}

\begin{figure*}[!htbp]
    \centering
    \includegraphics[width=\linewidth]{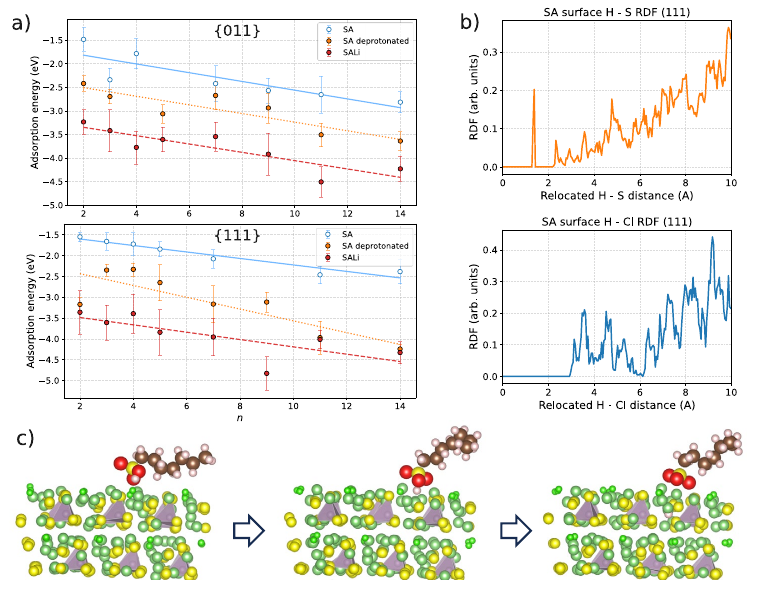}
    \caption{Finite-time deprotonation event observed for the sulfonic acid head group. (a) Adsorption energies of intact SA, deprotonated SA, and SALi on LPSCl $\{011\}$ and $\{111\}$. (b) RDFs between the detached H atom and surface S/Cl atoms. The detached H was mainly found near isolated sulfide-like S sites rather than S atoms within $\mathrm{PS_4}$ tetrahedra. (c) Representative snapshots of the deprotonation process for SA with $n=7$.}
    \label{fig:si_deprotonated}
\end{figure*}

H detachment was observed most clearly for the sulfonic acid head group (SA) within the finite MD trajectories. Fig.~\ref{fig:si_deprotonated}a compares intact SA, a trajectory classified as deprotonated SA, and fully prelithiated SALi on the LPSCl $\{011\}$ and $\{111\}$ surfaces. In these sampled configurations, the deprotonated case shows intermediate adsorption strength and is not equivalent to fully prelithiated SALi.

Within the inspected trajectories, detached H atoms were found mainly near isolated sulfide-like S sites rather than S atoms within $\mathrm{PS_4}$ tetrahedra (Fig.~\ref{fig:si_deprotonated}b). Fig.~\ref{fig:si_deprotonated}c shows one representative event for the $n=7$ SA molecule on LPSCl $\{111\}$. Because only a limited number of proton-transfer events were sampled, this analysis illustrates a possible pathway but does not provide a statistical deprotonation probability or establish the dominant experimental protonation state.

\section{Chain-length and coverage effects}

\subsection{Effect of molecular coverage on adsorption energy}
\label{sec:si_density_adsorption_energy}

\begin{figure}[!htbp]
    \centering
    \includegraphics[width=\linewidth]{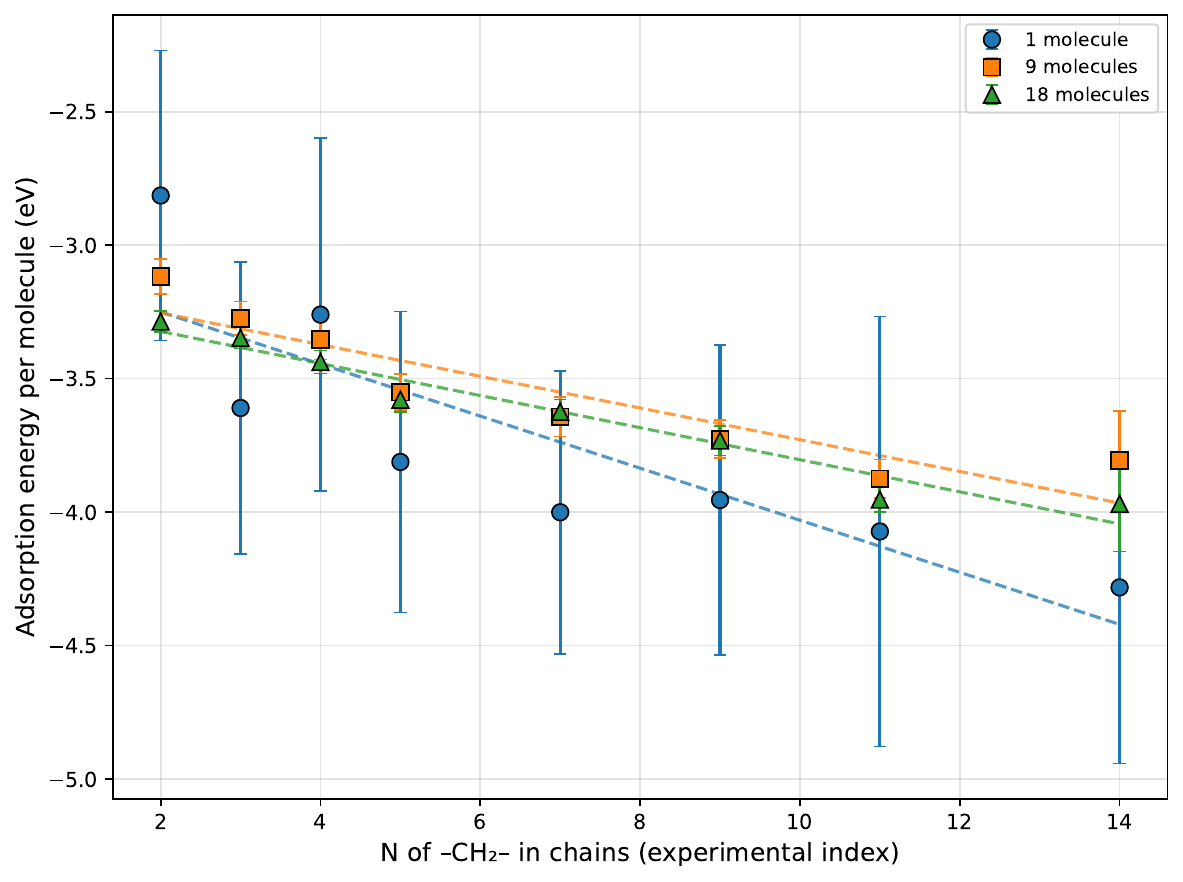}
    \caption{Adsorption energy per molecule of lithiated carboxylate chains on LPSCl $\{111\}$ as a function of alkyl-chain length at different molecular coverages. The similar chain-length trend across coverages indicates that increasing density mainly changes the balance between chain--surface and chain--chain interactions without substantially changing the head-group anchoring chemistry.}
    \label{fig:si_density_adsorption_energy}
\end{figure}

Fig.~\ref{fig:si_density_adsorption_energy} shows the adsorption energy per molecule of lithiated carboxylate chains on LPSCl $\{111\}$ at different coverages. Although the molecular conformation changes from flat-lying at low coverage to more upright at high coverage, the chain-length dependence remains similar. Increasing coverage therefore mainly shifts weak interactions from chain--surface to chain--chain contributions without substantially changing the head-group anchoring chemistry.

The larger error bars for the single-molecule systems reflect the broader range of adsorption sites and flat-lying configurations sampled by isolated molecules. At higher coverage, more sites are occupied simultaneously and the adsorption energy is averaged over several molecules, reducing the apparent variation.

\subsection{Coverage estimates and experimental additive loading}

To relate the simulated coverages to experimental additive loading, the LPSCl surface area was estimated by assuming spherical particles. Because the particle size and morphology are irregular, these values should be regarded as approximate. For particles of diameter $D$ and density $\rho=1.64$~g~cm$^{-3}$,
\begin{equation}
A_{\mathrm{LPSCl}} =
\frac{4 \pi \left(\frac{D}{2}\right)^2}
{\frac{4}{3} \pi \left(\frac{D}{2}\right)^3 \rho}
=\frac{6}{D \rho}.
\label{eq-ch2d:particle area}
\end{equation}

\begin{table*}[!htbp]
\centering
\begin{tabular}{lccc}
\hline
\textbf{Average particle size} & \textbf{Particle surface area} & \textbf{Specific surface area} & \textbf{mol\% for full coverage} \\
 & \textbf{$\mu$m$^2$} & \textbf{m$^2$\,g$^{-1}$} & \textbf{mol\%} \\
\hline
0.1~$\mu$m & 0.0314 & 36.6 & 8.2 \\
1~$\mu$m   & 3.14   & 3.7  & 0.8 \\
5~$\mu$m   & 78.5   & 0.73 & 0.2 \\
Literature solid* &  & 2.8  & 0.6 \\
Literature liquid-phase synthesis* &  & 20.6 & 4.6 \\
\hline
\end{tabular}
\caption{Theoretical particle and specific surface areas as a function of average particle size. *BET surface areas reported in.~\cite{Microtrac_BT009R1}}
\label{tab: 2d_area}
\end{table*}

The molar fraction required for complete coverage was estimated as
\begin{equation}
A = \frac{A_{\mathrm{LPSCl}} \cdot MW_{\mathrm{LPSCl}}}
{A_{\mathrm{HG}} \cdot N_A},
\label{eq-ch2d:coverage}
\end{equation}
where $N_A$ is Avogadro's number and $MW_{\mathrm{LPSCl}}$ is the molecular weight of LPSCl. Eighteen molecules on a 3.56~nm$^2$ slab correspond to 0.20~nm$^2$ per molecule, consistent with reported carboxylic head-group areas of 0.19--0.21~nm$^2$.~\cite{Denton2019-COOHsurface} The surface coverage of the experimental samples was not measured, so these values only indicate the order of magnitude of the loading required for a monolayer. More accurate estimates would require direct BET measurements.

\subsection{Chain-length-dependent experimental characterization}

\begin{figure*}[!htbp]
    \centering
    \includegraphics[width=\textwidth]{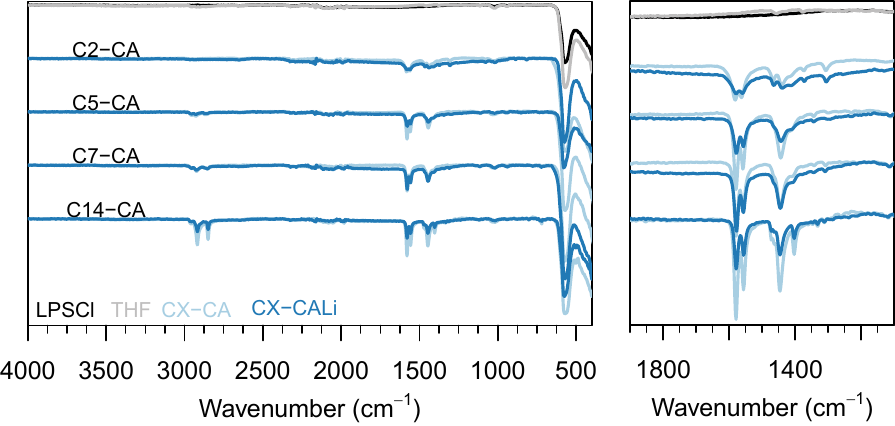}
    \caption{ATR--FTIR spectra of pristine additives and LPSCl coated with 10~mol\% carboxylic acids of different chain lengths.}
    \label{SI Fig: FTIR-Cx-CA}
\end{figure*}

The chain-length-dependent conductivity discussion in the main text is supported by the FTIR spectra in Fig.~\ref{SI Fig: FTIR-Cx-CA}, which compare the pristine carboxylic acids and the corresponding coated LPSCl samples across the investigated chain lengths.

The decrease in ionic conductivity with increasing additive loading, while chain length remains constant, supports the conclusion that the coating mildly hinders Li$^+$ transport.

\begin{figure*}[!htbp]
    \centering
    \includegraphics[width=0.4\textwidth]{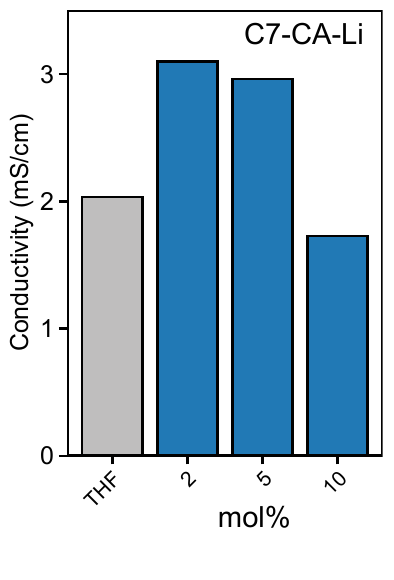}
    \caption{Ionic conductivity of LPSCl coated with different mol\% of C7-CALi.}
    \label{SI Fig: EIS-C7-CA}
\end{figure*} 

\begin{figure*}[!htbp]
    \centering
    \includegraphics[width=\textwidth]{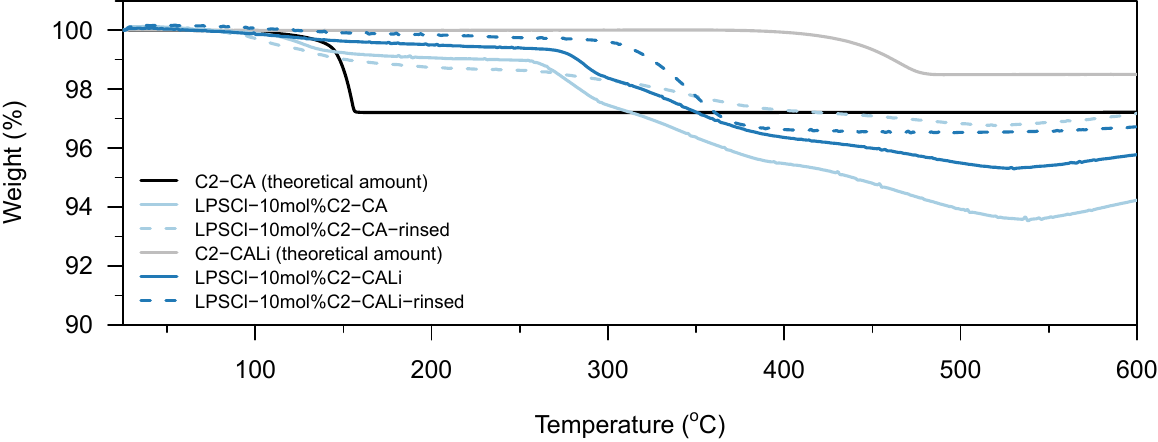}
    \caption{Thermogravimetric analysis between 25 and 600~\textdegree C for C2-CA and C2-CALi and 10~mol\%-coated LPSCl samples before and after THF washing and centrifugation.}
    \label{SI Fig:C2-CA}
\end{figure*}

The raw TGA traces for the chain-length series are provided below. Fig.~\ref{SI Fig:C2-CA} shows the short-chain C2 carboxylic-acid system.

\begin{figure*}[!htbp]
    \centering
    \includegraphics[width=\textwidth]{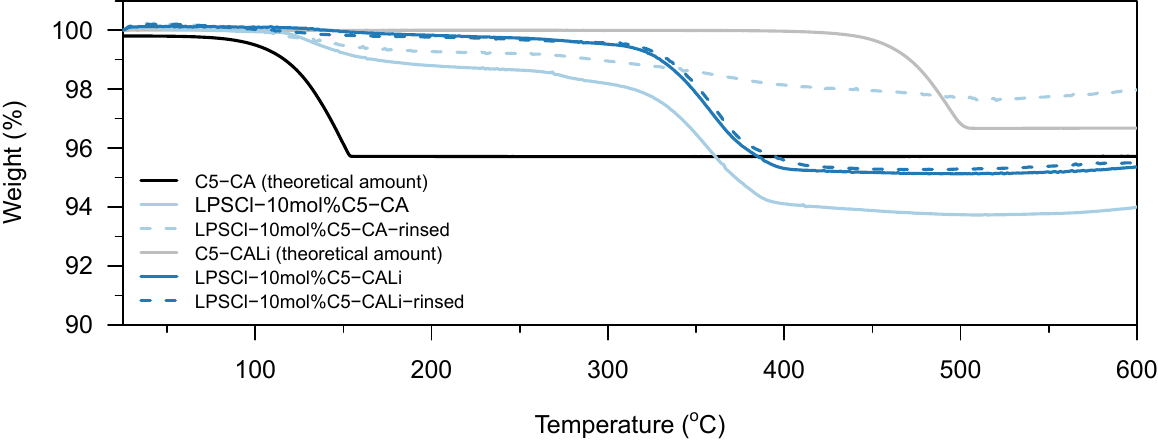}
    \caption{Thermogravimetric analysis between 25 and 600~\textdegree C for C5-CA and C5-CALi and 10~mol\%-coated LPSCl samples before and after THF washing and centrifugation.}
    \label{SI Fig:C5-CA}
\end{figure*}

Fig.~\ref{SI Fig:C5-CA} gives the corresponding intermediate-chain C5 data.

\begin{figure*}[!htbp]
    \centering
    \includegraphics[width=\textwidth]{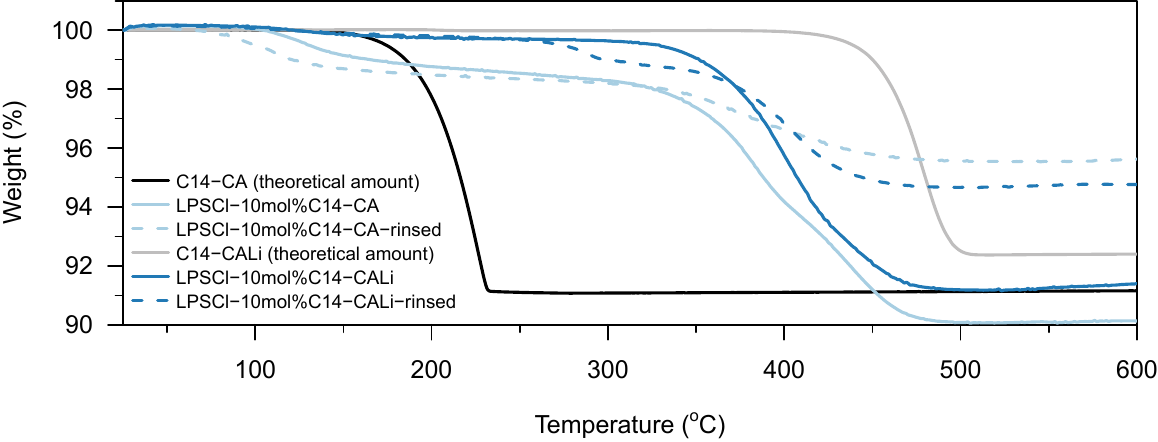}
    \caption{Thermogravimetric analysis between 25 and 600~\textdegree C for C14-CA and C14-CALi and 10~mol\%-coated LPSCl samples before and after THF washing and centrifugation.}
    \label{SI Fig:C14-CA}
\end{figure*}

The long-chain C14 carboxylic-acid system is shown in Fig.~\ref{SI Fig:C14-CA}.

\clearpage
\section{Moisture protection}

\subsection{Water--Li radial distribution analysis}
\label{sec:si_water_oli_rdf}

O$_\mathrm{water}$--Li RDFs were calculated for LPSCl $\{111\}$ surfaces coated with lithiated carboxylates at coverages of 9 and 18 molecules. A first peak near $r\approx2$~\AA{} indicates direct short-range access of water oxygen atoms to Li-rich interfacial sites, whereas suppression of this peak indicates reduced water penetration.

\begin{figure*}[!htbp]
    \centering
    \includegraphics[width=0.55\linewidth]{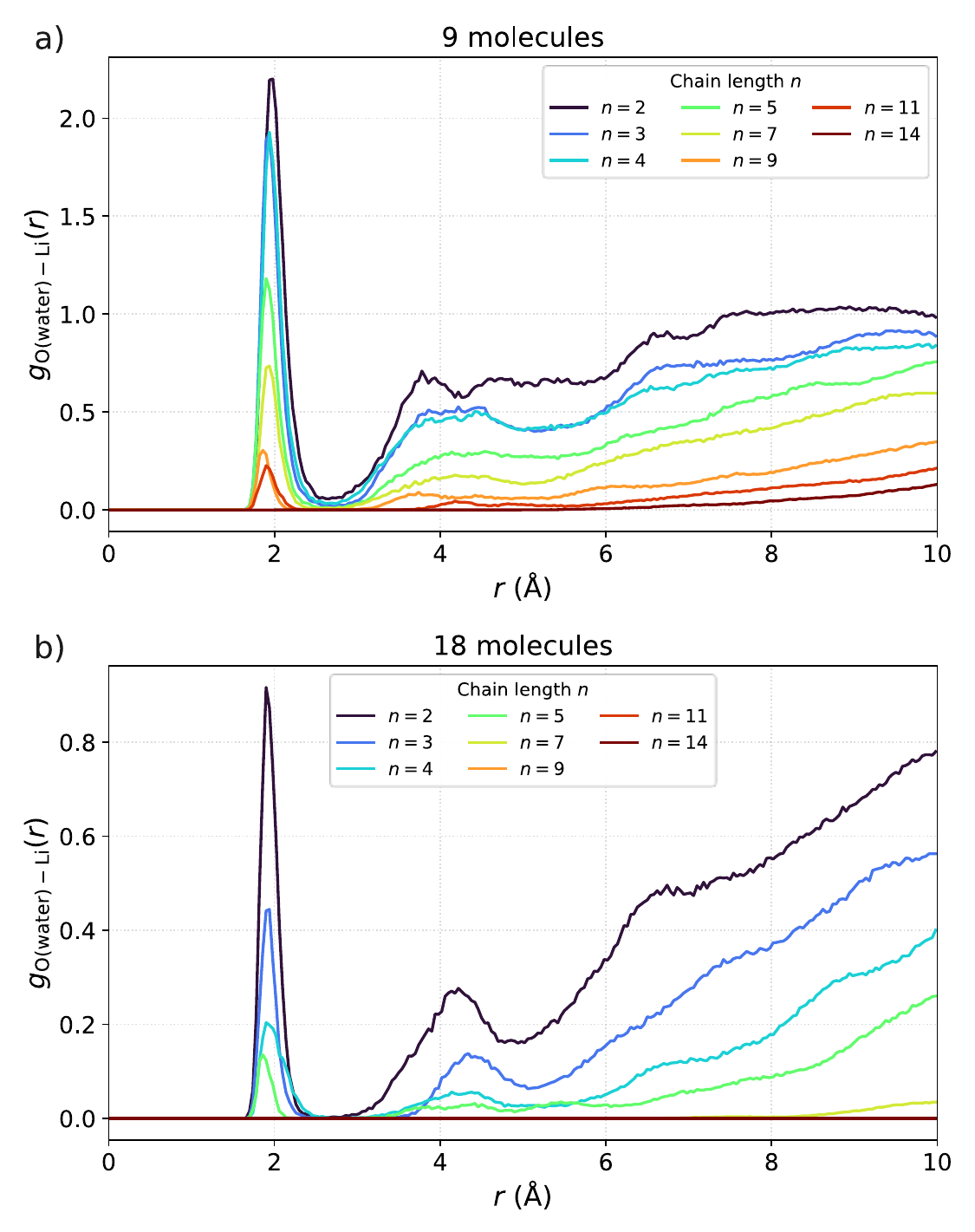}
    \caption{O$_\mathrm{water}$--Li radial distribution functions for LPSCl $\{111\}$ surfaces coated with lithiated carboxylate molecules of different alkyl-chain lengths. Panel (a) shows 9 adsorbed molecules and panel (b) 18 adsorbed molecules. Increasing chain length suppresses the first peak near $r\approx2$~\AA{}, and the suppression is stronger at higher coverage.}
    \label{fig:si_water_oli_rdf}
\end{figure*}

As shown in Fig.~\ref{fig:si_water_oli_rdf}, short-chain coatings, especially $n=2$ and $n=3$, retain pronounced first peaks. The peak decreases strongly with chain length and is nearly absent for $n=9$, $n=11$, and $n=14$. Suppression is stronger at 18-molecule coverage, consistent with a denser coating that more effectively limits direct water--Li contact.

\subsection{Water accessibility versus vertical chain extension}
\label{sec:si_height_water}

\begin{figure}[!htbp]
    \centering
    \includegraphics[width=0.6\linewidth]{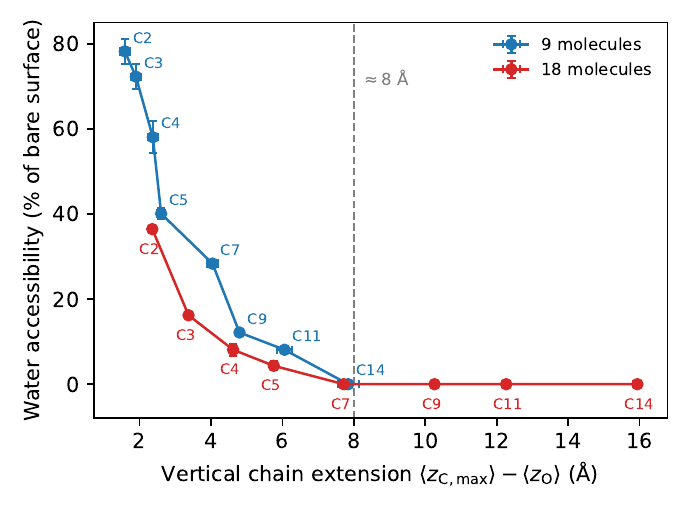}
    \caption{Water accessibility of lithiated carboxylate coatings on LPSCl $\{111\}$ as a function of the vertical chain extension $\langle z_{\mathrm{C,max}}\rangle - \langle z_{\mathrm{O}}\rangle$ at coverages of 9 (blue) and 18 (red) molecules. Labels give the chain length $n$. Error bars are standard deviations over the sampled MD frames. The dashed line marks an extension of 8~\AA{}.}
    \label{fig:si_height_water}
\end{figure}

To relate the water barrier to the height of the coating, Fig.~\ref{fig:si_height_water} shows the water accessibility (Fig.~\ref{fig:v2_moisture}) as a function of the vertical chain extension $\langle z_{\mathrm{C,max}}\rangle - \langle z_{\mathrm{O}}\rangle$ (Fig.~\ref{fig:ZCn}) at coverages of 9 and 18 molecules. At both coverages, the water accessibility decreases with increasing extension and vanishes at approximately 8~\AA{}, which is reached by $n=14$ at 9 molecules (7.8~\AA{}) and by $n=7$ at 18 molecules (7.7~\AA{}). Below this extension, the denser 18-molecule coating gives a lower water accessibility than the 9-molecule coating at a similar extension, indicating that the packing density also contributes to the barrier.

\subsection{Time-resolved H\texorpdfstring{$_2$S}{H2S} evolution during humid exposure}
\label{sec:si_h2s_evolution}

\begin{figure*}[!htbp]
    \centering
    \includegraphics[width=\linewidth]{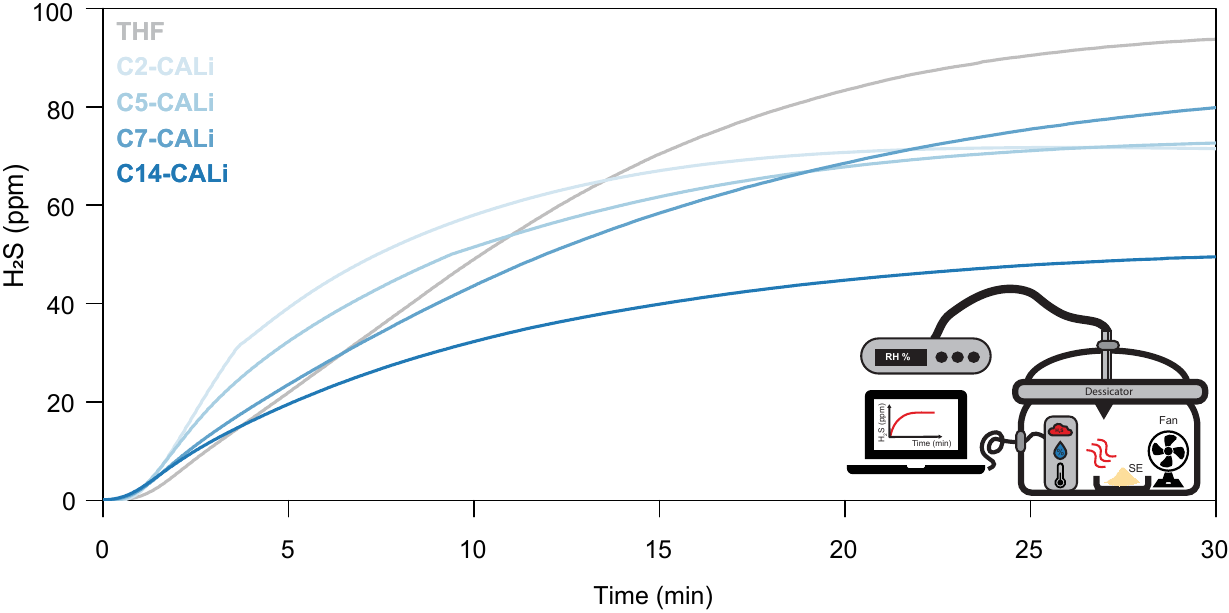}
    \caption{Time-resolved H$_2$S evolution during humid exposure of THF-treated and Cn-CALi-coated LPSCl powders at 20\% RH. The THF-treated reference shows the largest H$_2$S accumulation, whereas the molecular coatings reduce H$_2$S release. C14-CALi gives the lowest concentration over the measured time window. The initial slopes extracted from these traces are used in Fig.~\ref{fig:v2_moisture}.}
    \label{fig:si_h2s_evolution}
\end{figure*}

Fig.~\ref{fig:si_h2s_evolution} shows the time-resolved H$_2$S concentration during exposure of THF-treated and Cn-CALi-coated LPSCl powders to humid gas at 20\% RH. The THF-treated reference shows the largest H$_2$S accumulation, whereas all coated samples show reduced evolution. C14-CALi gives the lowest accumulation over the measured interval, consistent with improved moisture blocking by longer hydrophobic chains. The initial rates reported in the main text were extracted from these traces.

\end{document}